%% file: mqo.tex
\documentclass[11pt]{article}  %% The one col format
\usepackage{times}
\usepackage{defpaper}
\usepackage{psfig}
\usepackage{latexsym}
\renewcommand{\baselinestretch}{1.1}

\mergednumbering                        % Theorems, lemmas etc are numbered 

\eatreminders

\newcommand{\itemhead}[1]{\noindent{\bf #1:}\par}
\def\papernumber #1 raised #2 {
\vspace{-#2}
\vbox to 0pt{\hfill\framebox{\bf Paper Number #1}}
\vspace{#2}
}

\title{
Efficient and Extensible Algorithms for Multi Query Optimization}
\author{
{\bf Prasan Roy}\\ 
I.I.T. Bombay
\and
{\bf S. Seshadri}\\ % \footnote{Work done while at IIT Bombay}\\
Bell Labs.\ \\
\and
{\bf S. Sudarshan}\\
I.I.T. Bombay
\and
{\bf Siddhesh Bhobe} \\ % \thanks{Work done while at IIT Bombay}\\
PSPL Ltd. Pune
\and
\{prasan,sudarsha\}@cse.iitb.ernet.in\\
seshadri@research.bell-labs.com, siddhesh@pspl.co.in  
}

\date{}

\begin{document}

\eat{
\titlepage

\vspace*{.5in}

\centerline{\Large\bf Efficient and Extensible Algorithms for Multi Query Optimization}

\vspace{5mm}

\vspace{5mm}

\noindent{Authors:} \\
{\bf Prasan Roy},   I.I.T. Bombay, prasan@cse.iitb.ernet.in \\
{\bf S. Seshadri}, Bell Labs, seshadri@research.bell-labs.com \\
{\bf S. Sudarshan}, I.I.T. Bombay, sudarsha@cse.iitb.ernet.in \\
{\bf Siddhesh Bhobe}, PSPL Ltd. Pune, siddhesh@pspl.co.in

\begin{verbatim}

     Track:  Research

     Contact Author:
          Prasan Roy
          Computer Science and Engg. Dept.
          I.I.T. Powai,
          Mumbai 400076
          INDIA
          
          email: prasan@cse.iitb.ernet.in
          tel:   +91 22 576 7743
          fax:   +91 22 579 4290
\end{verbatim}

\pagebreak
}

\maketitle

%% if ONECOL uncomment
{ \small \renewcommand{\baselinestretch}{1.1}
\begin{abstract}
Complex queries are becoming commonplace, with the growing use of decision 
support systems. These complex queries often have a lot of common 
sub-expressions, either within a single query, or across multiple such 
queries run as a batch.  Multi-query optimization aims at exploiting common 
sub-expressions to reduce evaluation cost.  Multi-query optimization has
hither-to been viewed as impractical, since earlier algorithms were exhaustive,
and explore a doubly exponential search space.  

In this paper we demonstrate that multi-query optimization using heuristics
is practical, and provides significant benefits.
We propose three cost-based heuristic algorithms: Volcano-SH and Volcano-RU, 
which are based on simple modifications to the Volcano search strategy, and
a greedy heuristic.  Our greedy heuristic incorporates novel 
optimizations that improve efficiency greatly.  Our algorithms are designed
to be easily added to existing optimizers.  We present a performance study 
comparing the algorithms, using workloads consisting of queries from 
the TPC-D benchmark.  The study shows that our algorithms provide 
significant benefits over traditional optimization, at a very acceptable 
overhead in optimization time.  
\end{abstract}
}

\input{intro.tex}
\input{dag.tex}

% \input{daggen.tex}
\input{volcano.tex}

\input{greedy.tex}
\input{nestedindex.tex}

\input{perf.tex}

\input{relwork.tex}

\input{conclusions.tex}

%%%%%%%---------------------------------------------------------------------

{\small 
\renewcommand{\baselinestretch}{1.0}
\bibliographystyle{alpha}
\bibliography{multiquery}
}

%%%%%%%---------------------------------------------------------------------

%\newpage

%\fullversion{
%\appendix
%
%\input{appendix.tex}
%}

\end{document}

%% file: intro.tex
\section{Introduction}
\label{sec:intro}

Complex queries are becoming commonplace, especially due to the advent of
automatic tools that help analyze information from large data
warehouses. These complex queries often have a lot of common 
sub-expressions since i) they  make extensive use of views
%\footnote{ The WITH clause of SQL-3 simplifies use of temporary views.}
which are referred to multiple times in the query and 
ii) many of them are correlated nested queries in which parts of the 
inner subquery may not depend on the outer query variables, thus forming a 
common sub-expression for repeated invocations of the inner query. 

The scope for finding common sub-expressions increases greatly
if we consider a set of queries executed as a batch. 
For example, SQL-3 stored procedures may invoke several queries, which can 
be executed as a batch.  Data analysis/reporting often requires a batch 
of queries to be executed.
The work of \cite{wisc:xml} on using relational databases for storing 
XML data, has found that queries on XML data, written in a language 
such as XML-QL~\cite{xmlql}, need to be translated into a sequence 
of relational queries.

%%%%%%%%%%%%%%%
%Automatic tools often generate  a sequence of 
%SQL queries in response to a query in their native language. 
%Further, complex queries are often broken into a series of SQL queries.
%Such query sequences may have significant amounts of common 
%sub-expressions between queries.
%%%%%%%%%%%%%%%

The task of updating a set of related materialized views
also generates related queries with common sub-expressions
\cite{rss96:matview}.
% , which can benefit from multiquery optimization 
Materialized views are increasingly being supported by 
commercial database systems, and are used to speed up query processing.
In particular, data warehouses, which store large volumes of data, 
depend on materialized aggregate views for efficient query processing,
and the materialized views as well as expressions to incrementally
update them tend to have significant amounts of common subexpressions.
Sharing of common subexpressions is also of importance when the
expressions access remote data, and are therefore 
expensive \cite{shivku98:transview,levy99:opt}.
  
In this paper, we address the  problem of optimizing sets of queries
which may have common sub-expressions; this problem is referred to as 
{\em multi-query optimization}. 
We note here that common subexpressions are possible even 
{\em within} a single query; the techniques we develop deal with 
such intra-query common subexpressions as well.
%%%%%%%%%%%%%%
%The common sub-expression could be within the query, or
%across queries that are being concurrently optimized.
%%%%%%%%%%%%%%

Traditional query optimizers are not appropriate
for optimizing queries with common sub expressions, since
they make locally optimal choices, and may 
miss globally optimal plans as the following example
demonstrates. 
\begin{example}
\label{example:motivating}
Let $Q_1$ and $Q_2$ be two queries
whose locally optimal plans (i.e., individual best plans) are 
$(R \Join S) \Join P$ and $(R \Join T) \Join S$ respectively. 
The best plans for $Q_1$ and $Q_2$ do not have any
common sub-expressions.
However, if we choose the alternative plan $(R \Join S) \Join T$
(which may not be locally optimal) for $Q_2$,
then, it is clear that $R \Join S$ is a common sub-expression
and can be computed once and used in both queries.
This alternative with sharing of $R \Join S$ may be the globally 
optimal choice. 

On the other hand, blindly using a common
sub-expression may not always lead to a globally optimal strategy.
For example, there may be cases where the cost of joining the expression
$R \Join S$ with $T$ is very large compared to the cost
of the plan $(R \Join T) \Join S$; in such cases it may make no
sense to reuse $R \Join S$ even if it were available. 
\end{example}

Example~\ref{example:motivating} illustrates that the 
job of multi-query optimization, over and above that of
ordinary query optimization, is to 
(i) {\em recognize the possibilities of shared computation},
and (ii) {\em modify the optimizer search strategy to explicitly 
account for shared computation and find a globally optimal plan}.

While there has been work on multi-query optimization in the past
(\cite{tim:mul,kyu:imp,sg:tkde90,cls93:multi,joo:usi}),
prior work has concentrated primarily on exhaustive algorithms.
Other work has concentrated on finding common subexpressions as a 
post-phase to query optimization~\cite{fink82,shivku98:transview}, 
but this gives limited scope for cost improvement.
(We discuss related work in detail in Section~\ref{sec:related}.)
The search space for multi-query optimization is doubly exponential 
in the size of the queries, and exhaustive strategies are 
therefore impractical;
as a result, multi-query optimization was hitherto considered 
too expensive to be useful.

% Heuristics are therefore essential for efficient multi-query optimization.
% \reminder{Sesh: I am not sure it is correct to say search space
% is doubly exponential in the size of the query -- i think
% is ${2^n}^k$ where n is the size of the query and k is the 
% number of queries -- so i just made it doubly exponential
% and left it vague -- obviously I mean on two independent parameters}
% \reminder{Sud: this expression is nonsense with a single query, I've
% ignored your suggestion in the interest of making the paper believable!}

In this paper we show how to make multi-query optimization {\em practical},
by developing novel heuristic algorithms, and presenting a performance study 
that demonstrates their practical benefits. 

Our algorithms are based on 
an AND-OR DAG representation~\cite{rou82:view, rosenthal82,gra:vol}
to compactly represents alternative query plans.
The DAG representation ensures that they are {\em extensible}, in
that they can easily handle new operations and transformation rules.
The DAG can be constructed as in \cite{gra:vol,pel97:com}, with some
extensions to ensure that all common sub-expressions are detected and unified.
The DAG construction also takes into account sharing of computation based
on ``subsumption'' -- examples of such sharing include computing 
$\sigma_{A<5}(E)$ from the result of $\sigma_{A<10}(E)$.

\eat{
We present an algorithm for constructing the AND-OR DAG
based on algebraic transformations like in the Volcano optimization 
algorithm~\cite{gra:vol}, thereby making our approach {\em extensible}.
The novel aspect of our DAG construction algorithm lies in how it 
i) ensures that all common sub-expressions are detected and unified
and ii) takes into account sharing of computation based
on ``subsumption'' -- examples of such sharing include computing 
$\sigma_{A<5}(E)$ from the result of $\sigma_{A<10}(E)$.
}

%The DAG representation is outlined in Section~\ref{sec:dag:rep}.
%In this paper we consider how to add multi-query optimization to a
%query optimizer.  
%Our algorithms are based on algebraic transformations of queries, 
%as in the Volcano optimization algorithm \cite{gra:vol}, and are
%thereby very extensible.  Our algorithms, like 
%\cite{rou82:view, rosenthal82,gra:vol}, use an AND-OR 
%DAG representation to compactly represents alternative query plans.
%The DAG representation is outlined in Section~\ref{sec:dag:rep}.
%To address the first dimension outlined above, we present novel ways 
%to construct an AND-OR DAG representation of alternative query plans
%(Section~\ref{sec:daggen}).
%Our algorithms ensure that all common sub-expressions are detected and unified.
%The algorithms also take into account sharing of computation based
%on ``subsumption'' -- examples of such sharing include computing 
%$\sigma_{A<5}(E)$ from the result of $\sigma_{A<10}(E)$.

The task of the heuristic optimization algorithms is then to decide what
subexpressions should be materialized and shared. 
Two of the heuristics we present, Volcano-SH and Volcano-RU 
%(Sections~\ref{sec:volcano:sh} and \ref{sec:volcano:ru}) 
are lightweight modifications of the Volcano optimization algorithm.
%%%%%%
%and only consider subexpressions that are  part of the
%best plan as candidates for materializing and sharing. 
%%%%%%
The third heuristic is a greedy strategy which iteratively
picks the subexpression that gives the maximum benefit (reduction in cost)
if it is materialized and reused.  One of our important contributions here
lies in three novel optimizations of the greedy algorithm implementation,
that make it very efficient.  These are as follows:
\begin{enumerates}
\item We present a novel {\em incremental} algorithm for computing benefits 
of materializing different subexpressions.   

The motivation for incrementality is as follows.
The greedy strategy performs a large number of benefit computations,
with different sets of subexpressions chosen to be materialized.
Specifically, having picked a set $X$ of subexpressions to materialize,
to find the next one to materialize, the greedy strategy
must compute the benefit of every other subexpression $x$ that is
a candidate.

Having processed the earlier mentioned AND-OR DAG for finding
benefit of materializing subexpression $x_1$ (having already materialized
a set of subexpressions $X$), our incremental algorithm is able to 
compute the benefit of materializing a different subexpression $x_2$
incrementally, without revisiting the whole of the DAG.

\item 
The benefit of materializing a subexpression depends on what else is
materialized.  Suppose we computed the benefit of materializing
subexpressions $x_1, x_2, \ldots, x_n$, and expression $x_k$ was chosen
by greedy.  The next round of greedy has to recompute benefits of all
the remaining $x_i$'s, which can be expensive.

We present a monotonicity based technique for greatly reducing the number of
benefit recomputations required.  

\item We present an algorithm for computing sharability of subexpressions, 
allowing the greedy algorithm to be applied only to sharable nodes. 
\end{enumerates}
Our performance studies show that each of these optimizations leads to
a great improvement in the performance of the greedy algorithm.

\fullversion{
The benefit of materializing an expression also 
depends on the other expressions that have already been
chosen to be materialized.
}

In addition to choosing what intermediate expression results to materialize
and reuse, our optimization framework also chooses physical properties, 
such as sort order, for the materialized results.
Our algorithms also handle the choice of what
(temporary) indices to create on materialized results/database relations.
% Further, only indices that may be useful for future selects/joins are 
% considered.

Our algorithms can be easily extended to perform multi-query 
optimization on nested queries as well as multiple invocations of
parameterized queries (with different parameter values).
We also note that our algorithms can be made to work with System R
style bottom-up optimizers. 
% \reminder{Sud: New note above, add note on this in extensions/future 
% work section}

%Our techniques allow sharing of computation across
%different calls on the nested/parameterized query, as well
%as between a nested subquery and the outer level query.

% \end{enumerate}

\fullversion{
We now consider an example of how the optimizations we have proposed
can work together to significantly widen the applicability of multi-query
optimization.
The example uses a query based on the
TPC-D schema\footnote{This query is similar to 
%Example~\ref{example:motivating}. 
Query 2 of the TPC-D benchmark but we have modified both the query
and the schema slightly to conserve space.}.
\begin{example}
\label{example:motivating}
\vspace{-1mm}
{\em Find information about suppliers from India who carry at 
least 100 units of a part and who supply that part at 
the minimum cost amongst all big suppliers (those who carry 
at least 1000 units) from India of that part.} 
\begin{verbatim}
Schemas: part(p_partkey), supplier(s_suppkey, s_nation) 
         partsupp(ps_suppkey, ps_partkey, ps_availqty)

Query: select s_suppkey, p_partkey
       from part, supplier, partsupp
       where p_partkey = ps_partkey and s_suppkey = ps_suppkey 
       and ps_availqty > 100 and s_nation = "India"
       and ps_supplycost = 
               (select min(ps_supplycost)
                from  supplier, partsupp
                where p_partkey = ps_partkey and s_suppkey = ps_suppkey 
                and ps_availqty > 1000 and s_nation = "India")
\end{verbatim}
A possible execution strategy for the above complex query is to
first materialize the subexpression 
T1 : $\sigma_{ps\_availqty > 100 \mbox{ and } s\_nation = India} (supplier \Join partsupp)$, which
is a part of the outer query. 
Next, compute the expression  T2 :
$\sigma_{ps\_availqty > 1000} T1$ and then build an index on ps\_partkey
on  T2. The index will be used in the inner query. 
For each tuple of the outer query 
(obtained by joining the relations T1 and part), 
probe the index on the p\_partkey value of this tuple
to get the matching tuples for the inner query. 
Compute min(ps\_supplycost) for the matching tuples and check
if the outer tuple's ps\_supplycost is identical. This execution strategy may 
be the optimal strategy if the number of times the inner query is 
executed is very large and the optimal join order for the outer query is 
$(supplier \Join partsupp) \Join part$.
Notice that we have used a subsumption derivation,
created a temporary index
and identified sharing between the outer query and the inner
query in this example.
\end{example}
}

%Once again, Example~\ref{example:motivating}
%shows the benefits of doing so. This could also be
%useful in traditional optimizers since it allows us to consider
%using algorithms which need indexes (e.g., indexed nested loops join)
%on intermediate results also. We incorporate indexing into the Volcano
%framework as a physical property and this ensures that the
%search space does not explode.  Some commercial systems (e.g., Sybase)
%consider building indexes on intermediate results too.  
%Thus, in contrast to earlier literature
%we consider a rich set of possibilities for sharing.

%Our basic top-down optimizer also
%incorporates extensibility features of the EROC toolkit ~\cite{wil:ero},
%and the OPT++ optimization framework ~\cite{nav:opt}.
%Our optimizer is the first actual implementation of multi-query
%optimization that is integrated with a state-of-the-art top-down
%query optimizer based on the Volcano framework.
  
We have implemented a query optimizer based on the Volcano
optimization framework, and modified it to implement all our multi-query
optimization algorithms, at an additional effort of 
around 3500 lines of code.

We conducted a performance study of our multi-query optimization
algorithms, using queries from the TPC-D benchmark as well as
other queries based on the TPC-D schema.
Our study demonstrates not only savings based on estimated cost,
but also significant improvements in actual run times on a 
commercial database.

Our performance results 
show that our multi-query optimization algorithms give
significant benefits over single query optimization, at 
an acceptable extra optimization time cost.
The extra optimization time is more than compensated by the execution
time savings.
%Interestingly we found in some cases multi-query optimization actually
%reduces the optimization time, since it is able to reuse computation
%performed for earlier queries.
All three heuristics beat the basic Volcano algorithm, but in general
greedy produced the best plans, followed by Volcano-RU and Volcano-SH.

We believe that in addition to our technical contributions,
another of our contributions lies in showing how to engineer a 
practical multi-query optimization system --- one which can smoothly 
integrate extensions, such as indexes and nested queries, 
allowing them to work together seamlessly. 
In summer '99, our algorithms were partially prototyped on the 
Microsoft SQL Server optimizer, and multi-query optimization is 
currently being evaluated by Microsoft for possible inclusion in SQL Server.

%Our performance study also brings out the benefits due to various
%optimizations we have introduced.

\fullversion{
In particular, we show that unification reduces the sizes of
the DAGs generated, which in itself leads to time savings.
Our local pruning techniques further restrict the size of the 
DAG.  The optimizations of Greedy using monotonicity and incremental
update propagation are very effective, and as a result the cost of
the greedy heuristic is linear in the size of the DAG.
The addition of support for considering dynamic index creation, 
and subsumption improved performance, without significantly increasing 
the size of the DAGs.
% \reminder{IMP:
% Would be nice to show some numbers for index creation/subsumption
% paying off}
}

% \eat{
The rest of the paper is structured as follows:
We describe how to set up the search space for multi-query optimization in 
Section~\ref{sec:dag:rep}.  The Volcano-SH and the Volcano-RU 
heuristics are described in Section~\ref{sec:reusebased:mqo}. The greedy 
algorithm is described in Section~\ref{sec:greedyalgo}. Our extensions to 
create indexes on intermediate relations and nested queries are discussed
in Sections~\ref{sec:extnew}.  We describe the results of our performance 
study in Section~\ref{sec:perf}. 
Section~\ref{sec:related} discusses related work, while
Section~\ref{sec:future} discusses directions for future work.
We present our conclusions in Section~\ref{sec:conclusions}.

% }

%% file: dag.tex
\sections{Setting Up The Search Space For Multi-Query Optimization}
\label{sec:dag:rep}

As we mentioned in Section ~\ref{sec:intro}, the job of a 
multi-query optimizer is to 
(i) recognize possibilities of shared computation (thus essentially 
setting up the search space by identifying common sub-expressions) and 
(ii) modify the optimizer search strategy to explicitly account for 
shared computation and find a globally optimal plan. 
Both of the above tasks are important and crucial for a multi-query 
optimizer but are {\em orthogonal}. 
In other words, the details of the search strategy do not depend on 
how aggressively we identify common sub-expressions (of course, 
the efficacy of the strategy does). 
We have explored both the above tasks in detail, but choose 
to emphasize the search strategy component of our work in this paper, 
for lack of space. 
However, we outline the high level ideas and the intuition behind 
our algorithms for identifying common sub-expresions in this section 
and refer to the full version of the paper [RSSB98] for details 
at the appropriate locations in this section.

Before we describe our algorithms for identifying common-sub expressions, 
we describe the AND-OR DAG representation of queries. 
An AND--OR DAG is a directed acyclic graph whose nodes can be divided
into AND-nodes and OR-nodes;  the AND-nodes have only OR-nodes 
as children and OR-nodes have only AND-nodes as children. 
\fullversion{
The AND--OR DAG representation of queries can be viewed at two levels:
the {\em logical level} and the {\em physical level}.
We outline the two levels of representation below.
}

%We shall assume that queries are represented in algebraic form.
%For concreteness, we will be using the extended relational
%algebra 
%%including the groupby operation 
%(see, e.g., \cite{sks97:dbconcepts}), 
%although our basic techniques are general enough to handle other algebras.

\eat{
To simplify the description of our algorithms, we shall assume in much  
of the paper that there is only a single query (with common
subexpressions) being optimized. 
This assumption {\em does not} compromise generality, since 
the case when there are multiple queries can be easily reduced 
to the above case by introducing a zero-cost pseudo operation at the top, 
which has all the queries as its inputs.
}

% This representation is used, for example, in the Volcano optimizer generator,
% and provides a very compact and efficient way of representing alternative
% query processing strategies, without redundancy.

% The first level is the {\em logical level}, which represents
% alternative algebraic expressions that can be used to compute the query.
% The set of alternatives is defined by the logical search space generated 
% by a set of algebraic transformation rules provided by the optimizer 
% implementer.
% the search space that is generated by mapping the plans logical search 
% space to the physical (algorithm level) plans.

%\subsection{Logical AND-OR DAG}
%\label{ssec:log:dag}

\eat{
The AND-nodes represent logical operations, such as the join operation.
as well as specific implementations of the operations, 
such as merge-join or nested-loop join.
The OR-nodes represent a set of logically equivalent expressions.
%Each relation is considered to be an expression, and is thus represented by 
%an OR-node. 

Since OR-nodes in our DAG refer to a set of equivalent expressions, we shall
refer to them as {\em equivalence nodes} henceforth.
We distinguish two type of AND nodes in the DAG.
The first are the {\em operation nodes}, which represent logical
operations, for example the join of two inputs.
The second are the {\em operation nodes} which represent physical operations, 
namely operations with specific implementations, for example a merge-join or
nested loop join of two inputs.
}

An AND-node in the AND-OR DAG corresponds to an algebraic operation, 
such as the join operation ($\Join$) or a select operation ($\sigma$).
It represents the expression defined by the operation and its inputs. 
Hereafter, we refer to the AND-nodes as {\em operation nodes}.
An OR-node in the AND-OR DAG represents a set of logical 
expressions that generate the same result set; the set of such expressions
is defined by the children AND nodes of the OR node, and their inputs.
We shall refer to the OR-nodes as {\em equivalence nodes} henceforth.

\begin{figure*}
\centerline{
   % \mbox{\pdfimage width 6.0in {dagex.pdf} \relax}
   \psfig{file=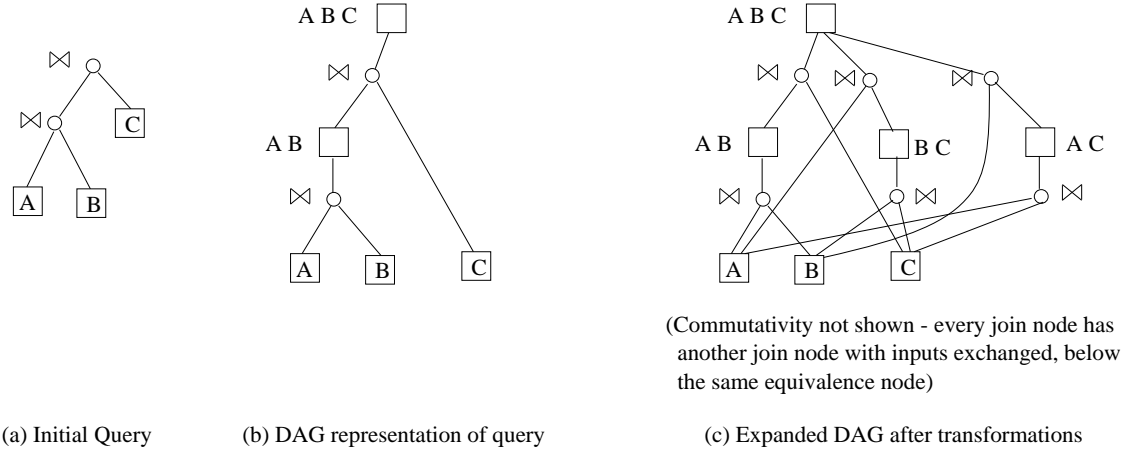,width=6in}
}
\caption{Initial Query and DAG Representations}
\label{fig:vol:dag}
\end{figure*}

The given query tree is initially represented directly in the AND-OR DAG 
formulation.
For example, the query tree of Figure~\ref{fig:vol:dag}(a) is initially
represented in the AND-OR DAG formulation, as shown 
in Figure~\ref{fig:vol:dag}(b).
Equivalence nodes (OR-nodes) are shown as boxes, while
operation nodes (AND-nodes) are shown as circles.

The initial AND-OR DAG is then expanded by applying all possible 
transformations on every node of the initial query DAG representing
the given set of queries. 
%For instance, consider the initial query $(A \Join B) \Join C$.
Suppose the only transformations possible are join associativity
and commutativity.
Then the plans $A \Join (B \Join C)$ and $(A \Join C) \Join B$,
as well as several plans equivalent to these modulo commutativity
can be obtained by transformations on the initial AND-OR-DAG
of Figure~\ref{fig:vol:dag}(b).
These are represented in the DAG shown in Figure~\ref{fig:vol:dag}(c).
We shall refer to the DAG after all transformations have been applied
as the {\em expanded DAG}.
%Figure~\ref{fig:vol:dag}(c) shows the expanded DAG after the 
%above transformations have been carried out.  
Note that the expanded DAG has exactly one equivalence node for every 
subset of $\{A, B, C\}$; the node represents all ways of computing 
the joins of the relations in that subset.  
\fullversion{
If we add another join with a relation $D$ to the query, 
the several plans where $D$ is joined with different expressions
for computing $A \Join B \Join C$ are all compactly represented by 
a single logical operation node representing a join, 
with $D$ as one input, and the logical equivalence node representing 
$A \Join B \Join C$ as the other input.
}
 
\eat{
Figure~\ref{fig:vol:dag} shows only operation nodes. In the
actual expanded DAG, an operation node is expanded into 
many physical operation nodes (each physical operation node representing
a  different implementation algorithm for performing the operation).
In addition, below each equivalence node representing a relation, 
we add physical operation nodes corresponding to relation scans.
}

% The OR--nodes represent equivalence classes. 
% An equivalence class represents the set of all operator 
% sub--DAGs that return the same bag of tuples. In other words, 
% it represents the set of all operator sub--DAGs such that any
% sub--DAG can be obtained from any other sub--DAG by applying 
% the transformation rules repeatedly. 

% \reminder{Sud: New stuff, check if we should keep.}
% We note that the System R bottom-up algorithm for join order optimization 
% can be viewed as constructing an AND-OR DAG for join orders.
% Equivalence nodes in the DAG correspond to sets of relations (for which 
% costs and best plans are maintained by System R), while operation nodes 
% are implicit.

We use the techniques of \cite{gra:vol} to efficiently generate the AND-OR
DAG representing the query.
We also use the optimizations of \cite{pel97:com}, which 
avoid duplicate derivations due to join associativity/commutativity. 
As shown in \cite{pel97:com}, the time complexity of top-down DAG 
generation for the case of joins is then the same as for bottom-up 
join-order enumeration as in System R style optimizers.
For lack of space we omit details of DAG generation; details may be 
found in \cite{rssb:tr}.

% Note also that DAG generation is not central to this paper; we
% concentrate here on the heuristic optimization algorithms.

\fullversion{
{\em For the benefit of referees, DAG generation is outlined in
Appendix~\ref{app:daggen}.}
}

\subsections{Extensions to DAG Generation For Multi-Query Optimization}
\label{sec:daggen}

\eat{
The initial logical DAG is expanded by applying all possible 
transformations on every node of the initial query DAG representing
the given set of queries.
The result is a single combined DAG, where common subexpressions are shared
between queries.
Our DAG generation algorithm roughly follows that used in the
Volcano optimizer generator \cite{gra:vol}, but there are some
important differences.  We outline the differences below;
full details may be found in \cite{rssb:tr}.
}

To apply multi-query optimization to a batch of queries, 
the queries are represented together in a single DAG, sharing 
subexpressions.  To make the DAG rooted, a pseudo operation 
node is created, which does nothing, but has the root equivalence nodes 
of all the queries as its inputs.

We now outline two extensions to the DAG generation algorithm to 
aid multi-query optimization.  

The first extension deals with identification of common subexpressions. 
If a query contains two subexpressions that are logically
equivalent, but syntactically different, 
(e.g., $(A \Join B) \Join C$, and $A \Join (B \Join C)$)
the initial query DAG would contain two different equivalence nodes
representing the two subexpressions.  
We modify the Volcano DAG generation algorithm so that whenever
it finds nodes to be equivalent (after applying join associativity)
it {\em unifies} the nodes, replacing them by a single equivalence node.

The Volcano algorithm uses a hashing scheme to detect repeated derivations, 
and avoids creating duplicate equivalence nodes due to cyclic derivations
(e.g., expression $e1$ is transformed to $e2$, which is then transformed
back to $e1$).
Our modification additionally uses the hashing scheme to detect and unify
duplicate equivalence nodes that were either pre-existing or
got created by transformations from different expressions.
Details of unification may be found in \cite{rssb:tr}.

\fullversion{
{\em For the benefit of referees, the algorithm for unification is described in 
Appendix~\ref{app:unif}.}
}

The second extension is to detect and handle
{\em subsumption}.
For example, suppose two subexpressions $e1$: $\sigma_{A<5}(E)$ and
$e2$: $\sigma_{A<10}(E)$ appear in the query.
The result of $e1$ can be obtained from the result of $e2$ by
an additional selection, i.e.,
$\sigma_{A<5}(E) \equiv \sigma_{A<5}(\sigma_{A<10}(E))$.
To represent this possibility we add an extra operation node 
$\sigma_{A<5}$ in the DAG, between $e1$ and $e2$.
% \fullversion{
Similarly, given $e3$: $\sigma_{A=5}(E)$ and $e4$: $ \sigma_{A=10}(E)$,
we can introduce a new equivalence node $e5$: $\sigma_{A=5 \vee A=10}(E)$ and 
add new derivations of $e3$ and $e4$ from $e5$.
The new node represents the sharing of accesses between the two selection.
In general, given a number of selections on an expression $E$, we 
create a single new node representing the disjunction of all the
selection conditions.
% }
Similar derivations also help with aggregations.
For example, if we have $e6$: $_{dno} {\cal G}_{sum(Sal)} (E) $ and
$e7$: $_{age} {\cal G}_{sum(Sal)} (E) $, we can introduce a new equivalence
node $e8$: $_{dno,age} {\cal G}_{sum(Sal)} (E) $ and add derivations of
$e6$ and $e7$ from equivalence node $e8$ by further groupbys on $dno$ and 
$age$.

% Our DAG generation algorithm introduces new equivalence nodes and operations 
% as above. Considering all subsumption possibilities leads to
% combinatorial explosion in the number of such nodes.
% Therefore we
% heuristically limit the possibilities considered. 
% \eat{
%   % but takes care to avoid combinatorial explosions in the number 
% of such nodes.
% }

The idea of applying an operation (such as $\sigma_{A<5}$ on one
subexpression to generate another has been proposed earlier
\cite{rou82:view,tim:mul,shivku98:transview}.
Integrating such options into the AND-OR DAG, as we do, clearly separates
the space of alternative plans (represented by the DAG) from the optimization 
algorithms.   Thereby, it simplifies our optimization algorithms,
allowing them to avoid dealing explicitly with such derivations. 

\eat{
Third, we implement the optimized transformation rules with marking of use
idea of \cite{pel97:com} which eliminates duplicate derivations of the same
expression due to join associativity transformations.
Unlike Volcano, our DAG expansion algorithm also expands the inputs to 
an operation completely, before applying transformations to the operation.
Thereby we are able to ensure that the set of transformations are 
completely applied, without having to retry the applicability of 
transformations (such as join associativity) which involve more than
one equivalence node.
}

% \{
% There are some interesting details here regarding {\em physical subsumption},
% where for e.g., $E$ sorted on $A$ can be generated from $E$ sorted on $B$.
% Details may be found in the full version of this paper \cite{rssb:tr} 
% }

% ({\em For the benefit of referees, our basic logical DAG generation
% algorithm is outlined in Appendix~\ref{app:daggen}.}
% \reminder{Need to add bits in the appendix algo re. where subsumption 
% is applied.  Perhaps also a bit on the hashing scheme can be provided there}
% )

\subsections{Physical AND-OR DAG}
\label{ssec:phys:dag}

Properties of the results of an expression, such as sort order,
that do not form part of the logical data model are
called {\em physical properties} \cite{gra:vol}.
Physical properties of intermediate results are important,
since e.g. if an intermediate result is sorted on a join attribute,
the join cost can potentially be reduced by using a merge join. 
This also holds true of intermediate results that are materialized and
shared.

It is straightforward to refine the above AND-OR DAG representation to
represent physical properties and obtain a physical AND-OR DAG.
\footnote{For example, an equivalence node is refined to
multiple physical equivalence nodes, one per required physical property,
in the physical AND-OR DAG.}
Our search algorithms can be easily understood on the above
AND-OR DAG representation (without physical properties), 
although they actually work on physical DAGs.  
Therefore, for brevity, we do not explicitly 
consider physical properties further; for details see \cite{rssb:tr}.
Our implementation indeed handles physical properties. 

\eat{
 Physical properties may 
either be requested by users (e.g., by using the SQL order-by clause) or 
may be useful for efficient evaluation using some algorithms, for instance, 
using the merge-join algorithm, which requires its input to be sorted.  
\fullversion{
The results of some algorithms may also have special physical properties: 
for instance, the result of merge-join is sorted on the join attributes. 
}
}

\eat{
To model optimization in the presence of physical properties, we 
use the notion of {\em physical AND-OR DAG}s.  Equivalence nodes in the
physical AND-OR DAG correspond to a logical expression coupled with 
a physical property.  Thus, logical equivalence node $A \Join B$ may
correspond to several physical equivalence nodes, sorted on different
attributes of $A \Join B$.  
Operation nodes in a physical AND-OR DAG correspond to algorithms,
such as hash-join or sort-merge join, or to ``physical property enforcers''
such as merge-sort.
}

\eat{
DAG generation as described in Section~\ref{sec:daggen} has one final 
stage, in which the logical AND-OR DAG is
expanded to a physical AND-OR DAG.  
DAG generation ensures that only those physical properties
that are useful for later operations or requested by users (corresponding
to ``interesting sort orders'' in System R) are generated.
}

\eat{
Logical and the physical AND-OR DAGs are quite symmetrical, and 
our search algorithms can be understood in the context of logical DAGs,
although they actually work on physical DAGs, thereby dealing cleanly 
with physical properties.
For brevity we omit further details of physical AND-OR DAGs.
}
%%%%%%%%%%%%%%%%%%%%%%%%%%%%%%%%%%%%%%%%%%%%%%%%%%%%%%%%%%%%%%%%%%%%%%%%%

% \reminder{ Add to perf.tex : belongs there .....
% We implement the transformation rules of \cite{pel97:com} 
% that keeps track of the derivation history 
% to eliminate duplicate derivations of the same
% expression due to join associativity transformations.
% }

%% file: volcano.tex
\eat{
\section{The Volcano Search Algorithm}
\label{ssec:volcano:search}

The Volcano Search Engine follows a top-down, goal-driven approach.
%combined with dynamic programming. 
It generates the logical DAG and expands it into the physical DAG on the fly.
We present below parts of the Volcano search algorithm, 
%rephrased in our terminology, 
that are relevant to our optimization algorithms.

%%%%%%%%%%%%%%%%%
% It allows the search space to be restricted to 
% only those subqueries that seem feasible and interesting, instead of the 
% entire space of possible moves.
%%%%%%%%%%%%%%%%%                 
\eat{
The top-down goal-driven approach allows (a) consideration of only 
subqueries that form part of a complete query, and 
(b) permits pruning of plans from consideration if while enumerating a 
subplan we find that the cost is already too high (higher than that of 
the best plan so far, or over a prespecified limit).
The dynamic programming approach ensures that once the best plan for an
equivalence node is found, it is stored and reused whenever that
equivalence node is revisited.
}
\fullversion{
The top-down goal-driven approach allows (a) consideration of only 
subqueries that form part of a complete query, and 
(b) permits pruning of plans from consideration if while enumerating a 
subplan we find that the cost is already too high (higher than that of 
the best plan so far, or over a prespecified limit).
The dynamic programming approach ensures that once the best plan for a
(physical) equivalence node is found, it is stored and reused whenever that
equivalence node is revisited.

The top-down approach (also known as backward chaining approach) 
is in contrast to the forward chaining strategy 
used in optimizers based on the System R approach to optimization, 
where smaller plans are enumerated and extended to create larger plans.
}

\eat{
{\em Subgoals} in Volcano are specified by a pair consisting of  
an equivalence node, representing the logical expression to be computed,
and a {\em physical property specification}\footnote{We use the
term physical property specification to refer to what is called
a physical property vector in Volcano.}, defining desired physical 
properties of the expression result.  For example if the outer level SQL query
had an {\sf order by} clause, the physical property desired would be
sorting on the specified columns.  As another example, the use of
a merge-join algorithm requires that the inputs be sorted on the join
attributes.  Thus, a physical operation node representing a merge-join would
require both its inputs to have the physical property of being sorted on
the join columns.
}

%%%%%%%%%%%%
% Attached with each equivalence node is a physical 
% property vector which describes some physical property
% of the output. Examples of physical properties are sort order, 
% partition vector etc. Certain operators (e.g., merge join) may
% require that its input have a certain physical property
% (e.g., sorted on the join attribute). 
% Failed attempts  at optimization of a given subtree may also be
% stored, since it can be used to terminate exploring that subtree in future.
% \subsubsection{The Search Algorithm}

Before the Volcano search algorithm is called on a query, the initial
query DAG corresponding to the given query is created.
Next, the initial DAG is fully expanded by applying the transformations
as described earlier, to get an expanded logical AND-OR DAG.
The search procedure is then called on the root of the expanded
logical AND-OR DAG.
(Note: the description of Volcano in \cite{gra:vol} does not
make this separation explicit, but the actual implementation does
follow this two phase approach for completeness of transformations\footnote{
Bill McKenna, Personal communication.}.)

The input to the Volcano search procedure is a logical equivalence node,
an initial physical property specification and an optional cost limit. 

\eat{
At this point the optimizer has three alternatives to explore. It can apply
a transformation rule, an implementation algorithm or an enforcer.
Applying a transformation rule results in a new expression which is
recursively optimized. An implementation algorithm is chosen if it
delivers the output according to the physical property specification supplied. 
In this case, the cost of the algorithm is added to the cost of
the inputs which is found by applying the optimization function
recursively on the inputs. Finally, an enforcer is one which enforces
a particular physical property specification, for example, a sort operator
permits considering hybrid hash join even if the final output is to be
sorted. After determining the possible alternatives, it orders these moves
according to the promise of these moves leading to the best plan.
It is necessary that this ordering be good, since it will lead to faster
pruning of the search space. Then, all alternatives are explored. 
While exploring the alternatives the minimum cost alternative so far 
can be used as the cost limit for the yet to be explored alternatives.
}

%The optimizer can then either
%exhaustively explore all possibilities or limit itself to a few good
%possibilities.  This decision is left to the implementor.
The search procedure tries alternative enforcers and algorithms for 
the operation nodes below the equivalence node, recursively calling 
itself to find the best plan for the inputs of the operation nodes.
A cost limit is passed as a parameter to the search algorithm, and 
if the cumulative cost of an operation node and the costs of the best 
plans for its inputs chosen so far exceeds the limit, the operation
node can be abandoned from consideration.

Once the best plan for an equivalence node, physical property pair
is found, it is stored in case it needs to be reused.
Therefore, in fact, the first thing to check 
before performing the above optimization for a given
node and a given physical property is to check for potential reuse.
If a plan matching the property specification is found among the plans
stored at the equivalence node,  and the plan satisfies the cost limit, 
the plan is returned; if a plan is found but does not satisfy the cost limit 
a failure indication is returned. 
If there is no plan for the expression and the property specification, 
then actual optimization (as described above) starts.

\fullversion{
The Volcano optimizer search algorithm does not take sharing of 
expressions into account.  Under this restriction, the principle 
of optimality holds: the best plan for an operation node must consist of
the best plans of its input equivalence nodes.
}

The best plan for a logical equivalence node, physical property pair
(thus, a physical equivalence node) is compactly specified by merely noting 
the corresponding physical operation node, and its input physical 
equivalence nodes.  The overall best plan is reconstructed when required
by recursively looking up the best plan for the inputs.

\fullversion{
The cost limit results in significant pruning, due to which large parts 
of the physical DAG that cannot form part of an optimal plan are 
not even generated.
If the result of an equivalence node is required to satisfy a physical
property, such as being sorted on attribute A, the Volcano search algorithm 
considers introducing an  enforcer algorithm, and recursively calls itself 
on the same logical equivalence node with no physical property requirement.

% Figure ~\ref{fig:sea} shows 
% a simplified version of the Volcano search algorithm.

The above description is a simplified version of the Volcano search
algorithm, and the algorithm as described in \cite{gra:vol}
is capable of supporting heuristic goal ordering and pruning as well.
}

% {\em For the benefit of the referees, we provide a detailed outline of our
% version of the Volcano search algorithm in Appendix~\ref{app:volcano}.}

\eat{
Figure ~\ref{fig:sea} shows the Modified Volcano Search Strategy.
\begin{figure}
\renewcommand{\baselinestretch}{1}
\ordinalg{
Function FindBestPlan(EqClass,PhyProp,CostLimit)\\
\\
if plan for EqClass is present \\
\> if cost of plan is less than CostLimit\\
\>\> return the plan as BestPlan\\
\> else if  cost of plan is more than CostLimit\\
\>\> return failure\\
else /* we need to optimize */ \\
\> if not Visited\\
\>\> apply all possible transformation rules to \\
\>\> operator subtrees of EqClass\\
\> create the set of possible moves from\\
\>\> applicable algorithms\\
\>\> applicable enforcers\\
\> order the set of moves by promise\\
\> while there are more moves to be applied do\\
\>\>if the move uses an algorithm\\
\>\>\> TotalCost = cost of algorithm\\
\>\>\> for each input of the algorithm while TotalCost\\
\>\>\> is less than CostLimit\\
\>\>\>\>determine physical property vector for input\\
\>\>\>\>call FindBestPlan for input subtree\\
\>\>\>\>add this cost to TotalCost\\
\>\>if the move uses an enforcer\\
\>\>\>find new physical property vector\\
\>\>\>call FindBestPlan for EqClass with new PhyProp\\
\>\>check if plan found is better than current best plan\\
\>\>and replace if necessary\\
\>return best plan found\\
}
\caption{The Modified Volcano Search Algorithm}
\label{fig:sea}
\end{figure}

%\begin{figure}
%\centerline{\psfig{file=algorithm.ps}}
%\caption{The Modified Volcano Search Algorithm}
%\label{fig:sea}
%\end{figure}
}
}
%%%%%%%%%%%%%%%%%%%%%%%%%%%%%%%%%%%%%%%%%%%%%%%%%%%%%%%%%%%%%%%%%%%%%%%%%%%%%%

\sections{Reuse Based Multi-Query Optimization Algorithms}
\label{sec:reusebased:mqo}

In this section we study a class of multi-query optimization
algorithms based on reusing results computed for other 
parts of the query. We present these as extensions of the Volcano
optimization algorithm.
Before we describe the extensions, in Section~\ref{ssec:volc:mat},
we very briefly outline the 
basic Volcano optimization algorithm, and how to extend it to find
best plans given some nodes in the DAG are materialized.
Sections~\ref{sec:volcano:sh} and \ref{sec:volcano:ru} then present 
two of our heuristic algorithms, Volcano-SH and Volcano-RU.

\subsection{Volcano Optimization Algorithm and Materialized Views}
\label{ssec:volc:mat}

The Volcano optimization
algorithm  operates on the expanded DAG generated earlier.
It finds the best plan for each node by performing a depth first
traversal of the DAG starting from that node as follows.
Costs are defined for operation and equivalence nodes.
The cost of an operation node $o$ is defined as follows:\\
\hspace*{1cm}
$cost(o) = $ cost of executing $(o)$ + $\Sigma_{e_i \in children(o)}
cost(e_i)$\\
The children of $o$ (if any) are equivalence nodes.\footnote{The 
cost of executing an operation $o$ also takes into account the cost 
of reading the inputs, if they are not pipelined.
}
The cost of an equivalence node $e$ is given as \\
\hspace*{1cm}
$cost(e) = min \{ cost(o_i) | o_i \in children(e) \} $\\
and is $0$ if the node has no children (i.e., it represents a relation).

% the cost of executing the operation plus the cost of its 
% children (i.e., inputs to the operator) while
% the cost of an equivalence node is the minimum of the costs of 
% its children (i.e., alternatives for obtaining the result represented
% by the equivalence node).  

Volcano also caches the best plan it finds for each equivalence node, 
in case the node is re-visited during the
depth first search of the DAG. A branch and bound pruning is also
performed by carrying around a cost limit; for simplicity,
we disregard pruning in this paper.
For lack of space we omit details, but refer readers to
\cite{gra:vol}.

\fullversion{
({\em For the benefit of referees, the Volcano optimization
algorithm is briefly outlined in Appendix~\ref{app:volcano}.})
}

Now we consider how to extend Volcano to find best plans, given
that (expressions corresponding to) some equivalence nodes in the 
DAG are materialized.  
Let $reusecost(m)$ denote the cost of reusing the materialized 
result of $m$, and let $M$ denote the set of materialized nodes.

To find the cost of a node given a set of nodes $M$ have been materialized, 
we simply use the Volcano cost formulae above, with the following change.
When computing the cost of a operation node $o$, if an input equivalence
node $e$ is materialized (i.e., in $M$), use the minimum of 
$reusecost(e)$ and $cost(e)$ when computing $cost(o)$.  
Thus, we use the following expression instead:\\
\hspace*{1cm}
$cost(o) = $ cost of executing $(o)$ + $\Sigma_{e_i \in children(o)}
  C(e_i)$\\
\hspace*{2cm}where $C(e_i) = cost(e_i)$ if $e_i \not\in M$, and 
	     $=min(cost(e_i), reusecost(e_i))$ if $e_i \in M$.\\

\fullversion{
The traditional Volcano search algorithm does not guarantee finding
optimal plans with common sub-expressions for two reasons.  First, costs
of shared sub plans are counted as many times as they occur.  Second,
Volcano only looks at locally optimal plans, whereas the global optimal
plan may contain sub-plans that are not locally optimal.

\eat{
We first describe the Cost Set Model, which avoids multiple counting of
common sub-expression cost, and an extension of Volcano using the Cost 
Set Model.
}

We first describe a very simple and inexpensive strategy for obtaining DAG
query plans from the best plan obtained by Volcano.  Next, we describe
a marking based extension of Volcano which avoids multiple counting of
some common sub-expressions.  We then describe algorithms that change the
search strategy.  We describe an exhaustive backtracking algorithm that
finds optimal plans and outline how to incorporate  memoization into the
exhaustive algorithm.  Finally, we describe a greedy heuristic based on
an alternative incremental marking of shared nodes based approach.
}

%%%%%%%%%%%
%% Thus, the plan for the initial part of the query is globally optimal,
% if the query were restricted to the initial part alone, but is not
% necessarily globally optimal for the whole query.  
%%%%%%%%%%%%

\subsections{The Volcano-SH Algorithm}
\label{sec:volcano:sh}

In our first strategy, which we call Volcano-SH, the expanded DAG
is first optimized using the basic Volcano optimization algorithm.
The best plan computed for the virtual root is the combination of
the Volcano best plans for each individual query.
The best plans produced by the Volcano optimization algorithm may have
common subexpressions.  Thus the consolidated best plan for the root of
the DAG may contain nodes with more than one parent, and 
is thus a DAG-structured plan.\footnote{
The ordering of queries does not affect the above plan.
}
The Volcano-SH algorithm works on the above consolidated best plan, 
and decides in a cost based manner which of the nodes to materialize 
and share.

Since materialization of a node involves storing
the result to the disk, and we assume pipelined execution of operators,
it may be possible for recomputation of a node to be cheaper than the
cost of materializing and reusing the node. In fact, in our experiments
in Section~\ref{sec:perf}, there were quite a few occasions when it was
cheaper to recompute an expression.  

Let us consider first a naive (and incomplete) solution.
Consider an equivalence node $e$.  
Let $cost(e)$ denote the computation cost of node $e$. 
Let $numuses(e)$ denote the number of times node $e$ is used in 
course of execution of the plan.
Let $matcost(e)$ denote the cost of materializing node $e$. 
As before, $reusecost(e)$ denote the cost of reusing the materialized 
result of $e$.
Then, we decide to materialize $e$ if 
\[ cost(e) + matcost(e) + reusecost(e) \times (numuses(e)-1) <
              numuses(e) \times cost(e)\]
The left hand side of this inequality gives the
cost of materializing the result when first computed, and using the
materialized result thereafter; the right hand side gives the cost of
the alternative wherein the result is not materialized but recomputed
on every use.  
The above test can be simplified to 
\begin{equation}
 matcost(e)/(numuses(e)-1)  + reusecost(e) < cost(e)
\label{eqn:matcond}
\end{equation}

% taken optimally.  Let its materialization cost be
% $C^M_e$, and the cost of using the materialized result be $C^R_e$. 

The problem with the above solution is that $numuses(e)$ and 
$cost(e)$ both depend on what other nodes have been materialized,  
For instance, suppose node $e_1$ is used twice in computing node $e_2$,
and node $e_2$ is used twice in computing node $e_3$.
Now, if no node is materialized, $e_1$ is used four times in computing
$e_3$.  If $e_2$ is materialized, $e_1$ gets used twice in
computing $e_2$, and $e_2$ gets computed only once.
Thus, materializing $e_2$ can reduce both $numuses(e_1)$ and 
$cost(e_3)$.
 
In general, $numuses(e)$ depends on which ancestors of $e$ in the 
Volcano best plan are materialized, and $cost(e)$ depends on
which descendants have been materialized.
Specifically, $numuses(e)$ can be computed 
recursively based on the number of uses of the parents of $e$:
$numuses(root) = 1$, while for all other nodes, 
$numuses(e) = \sum_{p \in parents(e)} U(p)$, where $U(p) = numuses(p)$
if $p$ is not materialized, and $= 1$ if $p$ is materialized.
Thus, computing $numuses$ requires us to know the materialization 
status of parents.
On the other hand, as we have seen earlier, $cost(e)$ depends on
what descendants have been materialized.

A naive exhaustive strategy to decide what nodes in the Volcano best plan to
materialize is to consider each subset of the nodes in the
best plan, and compute the cost of the best plan given that all 
nodes in this subset are materialized at their first computation; 
the subset giving the minimum cost is selected for actual 
materialization.  
Unfortunately, this strategy is exponential in the number of nodes in the Volcano
best plan, and therefore is very expensive; we require cheaper heuristics.

\begin{figure}[t]
\begin{small}
\ordinalg{
Procedure {\sc Volcano-SH}($P$) \\
{\em Input:} \> \> Consolidated Volcano best plan $P$ for virtual root of DAG \\
{\em Output:} \> \> Set of nodes to materialize $M$,
			and the corresponding best plan $P$ \\
{\em Global variable:}  $M$, the set of nodes chosen to be materialized \\
\> $M = \{~\} $ \\
\> Perform prepass on $P$ to introduce subsumption derivations \\
%%%%%%%%%
% \> Compute $cost^+(e)$, $numuses^+(e)$ and $numuses^-(e)$ for each node
%   in the plan $P$. \\
% \> Let $S = \{ e \in P | matcost(e)/(numuses^+(e)-1)  + reusecost(e) < 
% 			cost^+(e) \} $ \\
%%%%%%%%%
\> Let $C_{root} =$  {\sc ComputeMatSet}($root$) \\
\> Set $C_{root} = C_{root} + \sum_{d \in M} (cost(d)+matcost(d))$  \\
\> Undo all subsumption derivations on $P$ where the subsumption node is not
chosen to be materialized. \\
\> return (M,P) \\
\\
Procedure {\sc ComputeMatSet}$(e)$ \\
\> If $cost(e)$ is already memoized, return $cost(e)$  \\
\> Let operator $o_e$ be the child of $e$ in $P$ \\
\> For each input equivalence node $e_i$ of $o_e$ \\
\> \> Let $C_i$= {\sc ComputeMatSet($e_i$)} ~~~~ // returns computation cost of $e_i$ \\
\> \> If $e_i$ is materialized, let $C_i = reusecost(e_i)$  \\
\> Compute $cost(e)$ = cost of operation $o_e$ + $ \sum_{i}C_i $ \\
%%%%   + \sum_{d \in S^M_e} (cost(d)+matcost(d)$ 
%%%% \> \> // where $S^M_e =$ set of descendents of $e$ in $P$ already decided
%%%% to be materialized \\
%% \> \> // \> \> \> \> and having no ancestor in $P$ that is not a
%% 		descendent of $e$ \\
\> If ($matcost(e)/(numuses^-(e)-1)  + reusecost(e) < cost(e) $) \\
\> \> If $e$ is {\em not} introduced by a subsumption derivation  \\
\> \> \>  add $e$ to $M$ ~~~~~~// Decide to materialize $e$  \\
\> \> else if $ cost(e) + matcost(e) + reusecost(e)*(numuses^-(e)-1) $ is less 
          than \\
\> \> \> \> savings to parents of $e$ due to introducing materialized $e$\\
\> \> \>  add $e$ to $M$ ~~~~~~// Decide to materialize $e$  \\
\> Memoize and return $cost(e)$
}
\end{small}
\vspace{-2mm}
\caption{The Volcano-SH Algorithm}
\label{fig:volcano-sh}
\end{figure}

To avoid enumerating all sets as above, the Volcano-SH algorithm,
which is shown in Figure~\ref{fig:volcano-sh}, traverses the tree
bottom-up.  As each equivalence node $e$ is encountered in the traversal,
Volcano-SH decides whether or not to materialize $e$. 
When making a materialization decision for a node,
the materialization decisions for all descendants ia already known.
When Volcano-SH is examining a node $e$, let $M$ denote the set of
descendants of $e$ that have been chosen to be materialized.
Based on this, we can compute $cost(e)$ for a node $e$, 
as described in Section~\ref{ssec:volc:mat}. 

To make a materialization decision for a node $e$, we also need to 
know $numuses(e)$.  Unfortunately, $numuses(e)$ depends on the materialization
status of its parents, which is not fixed yet.
To solve this problem, the Volcano-SH algorithm uses an 
underestimate $numuses^-(e)$ of number of uses of $e$.
Such an underestimate can be obtained by simply counting
the number of ancestors of $e$ in the Volcano best plan.  
We use this underestimate in our cost formulae,
to make a conservative decision on materialization.\footnote{We
also developed and tried out a more sophisticated underestimate. 
We omit it from here for brevity, and because it only lead to
a minor improvement on performance.}

Based on the above, Volcano-SH makes the decision on materialization
as follows: node $e$ is materialized if 
\begin{equation}
 matcost(e)/(numuses^-(e)-1)  + reusecost(e) < cost(e) 
\label{eqn:sh-matcond}
\end{equation}
Note that here we use the lower bound $numuses^-(e)$ in place of $numuses(e)$. 
Using the lower bound guarantees that if we decide to materialize
a node, materialization will result in cost savings.

\reminder{May want an example here...}

The final step of Volcano-SH is to factor in the cost of computing and
materializing all nodes that were chosen to be materialized.
Thus, to the cost of the pseudoroot computed as above, we add
$\sum_{m \in M}(cost(m) + matcost(m))$, where $M$ is the set of 
nodes chosen to be materialized.

Let us now return to the first step of Volcano-SH.
Note that the basic Volcano optimization  algorithm will not exploit 
subsumption derivations, such as deriving $\sigma_{A<5}(E)$ by using
$\sigma_{A<5}(\sigma_{A<10}(E))$,
since the cost of the latter will be more than the former,
and thus will not be locally optimal.

To consider such plans, we perform a pre-pass, checking for subsumption 
amongst nodes in the plan produced by the basic Volcano optimization 
algorithm.  If a subsumption derivation is applicable, we replace
the original derivation by the subsumption derivation.
At the end of Volcano-SH, if the shared subexpression is not chosen
to be materialized, we replace the derivation by the original
expressions.  In the above example, in the prepass we replace
$\sigma_{A<5}(E)$ by $\sigma_{A<5}(\sigma_{A<10}(E))$.
If $\sigma_{A<10}(E)$ is not materialized, we replace 
$\sigma_{A<5}(\sigma_{A<10}(E))$ by $\sigma_{A<5}(E)$.

\eat{
% a cost-based decision is made on whether to use it in conjunction 
% with materialization of the shared expression node.
% For example, the cost of materializing $\sigma_{A<10}(E)$ and computing 
% $\sigma_{A<5}(E)$ from it is compared to the cost of computing
% each directly from $E$; if the former is cheaper, the shared expression 
% $\sigma_{A<10}(E)$ is materialized.
}

% Note that since the rest of the Volcano-SH algorithm requires 
% a single plan to be considered, 

\eat{
First, we find the common subexpressions by traversing the  
optimal plan in a depth first manner. During the traversal,
we also keep
track of the number of times an equivalence node is shared (number of
uses)  in the optimal plan. The choice of materialization is done by another 
depth-first traversal of common subexpressions.
If the materialization-cost + (number-of-reuses $\times$ reuse-cost)
is less than (number-of-reuses $\times$ cost-of-recomputation), a decision 
is made to materialize and reuse the common subexpression, else 
it is not materialized. The cost of recomputation is
calculated based on the choice of materialization of the
descendants (which is known at this point of time
since the traversal is depth-first).

Note that the basic Volcano optimization  algorithm will not exploit 
subsumption derivations, such as deriving $\sigma_{A<5}(E)$ by using
$\sigma_{A<5}(\sigma_{A<10}(E))$,
since the cost of the latter will be more than the former,
and thus will not be locally optimal.
Therefore, we perform a pre pass, before the main part above,
checking for subsumption amongst nodes in the
plan produced by the basic Volcano optimization algorithm. 
If a subsumption derivation is applicable, 
%amongst nodes in the locally optimal plans,
a cost-based decision is made on whether to use it in conjunction 
with materialization of the shared expression node.
For example, the cost of materializing $\sigma_{A<10}(E)$ and computing 
$\sigma_{A<5}(E)$ from it is compared to the cost of computing
each directly from $E$; if the former is cheaper, the shared expression 
$\sigma_{A<10}(E)$ is materialized.
%\footnote{$\sigma_{A<10}(E)$ could also 
%be a common subexpression and therefore, the decision to 
%materialize  $\sigma_{A<10}(E)$ or not could 
%be made in conjunction with the decision for common subexpressions
%rather than in two independent phases as we have outlined.}.

The above techniques for choosing nodes to materialize is a heuristic 
since the decision to materialize or not is taken locally,
assuming all shared ancestors are materialized.
If a shared ancestor node is not materialized, then the actual number 
of uses of a node may be greater than that computed by the first depth 
first traversal. 
A more expensive technique that eliminates the order dependence would 
be to consider materializing each subset of the set of common subexpressions 
and choose the subset with the smallest cost.  However, even this would still
have the problem addressed by Volcano-RU in the next section.
}

\eat{
The effectiveness of the above algorithm depends on how tightly 
$numuses^-(e)$ bounds $numuses(e)$ from below. 
We can improve the algorithm by obtaining a tighter lower bound as follows.
The key observation is to note that materializing an additional node 
in the plan can only decrease the number of uses of any node in the plan. 
Thus, if we can compute a superset (called $M^+$ henceforth) of the set $M$
of nodes chosen for  materialization, then we can take $ numuses^-(e)$ 
as the number of uses of the node $e$ assuming that all nodes in $M^+$ 
are materialized. We show how to compute such an $M^+$ next.

Let $numuses^+(e)$ and $cost^+(e)$ be the number of uses and cost 
respectively of node $e$ given that {\em no} node in the plan is 
materialized.  
Now, since $numuses^+(e) \geq numuses(e)$ and $cost^+(e) \geq cost(e)$, 
we need only consider for potential materialization the equivalence 
nodes $e$ in the plan that satisfy
\begin{equation}
 matcost(e)/(numuses^+(e)-1)  + reusecost(e) < cost^+(e)
\label{eqn:matcond-filter}
\end{equation} 
since the remaining nodes  cannot satisfy the inequality (\ref{eqn:matcond}), 
and hence cannot be in $M$. 
Thus, we take $M^+$ as the set of nodes in the plan that satisfy the 
inequality (\ref{eqn:matcond-filter}).
Given this superset $M^+$ of the set of nodes that can get materialized,
we can compute the number of uses of each node.
This gives us a tighter underestimate for $numuses^-(e)$.

Our implementation of Volcano-SH is based on the above improvement. 
In our performance study (Section~\ref{sec:perf}),
we found that the performance overheads due to the additional steps 
required as above in the improved algorithm were a negligible fraction 
of the overall optimization cost.
}

The algorithm of \cite{shivku98:transview} also finds best plans and
then chooses which shared subexpressions to materialize.
Unlike Volcano-SH, it does not factor earlier materialization choices
into the cost of computation.

%\footnote{We did
%not observe any difference in our experiments in the cost of the
%plan produced by these algorithms.}.
%However, using a lower degree of sharing is conservative in that it never
%overestimates the benefits of materializing a node.

\fullversion{
\reminder{
Future work: 
Implement the above after incorporating materialization costs, 
as well as pipelining alternatives into our optimizer.
}
}
\eat{
The above strategy is an extension of the well known idea of
common subexpression elimination, which takes recomputation costs 
into account.}
% Sesh: not sure this is very well known -- while we should
% not claim it is novel -- we should not do this to ourelves.
% Unless we can quote references and say exactly how we differ
% I think harmless enough to shut up.

% Note that, if an equivalence node 
% $e1$ is made a parent of a new logical operation node (whose
% child is $e2$) because of a subsumption derivation, then
% the new logical operation node can never
% be part of a locally optimal plan chosen by Volcano.

% For instance it may be cheaper to perform the selection $\sigma_{A < 5} (E)$
% directly on $E$, if $E$ had an index on $A$,
% than computing $\sigma_{A < 5}$ on the materialized result
% of (\sigma_{A < 10} (E))$.

\subsections{The Volcano-RU Algorithm}
\label{sec:volcano:ru}

\fullversion{
Volcano-SH considers only nodes in the best plan produced by the
basic Volcano optimization algorithm as candidates 
for materializing and sharing. The basic Volcano optimization
algorithm, however, makes locally optimal choices at every node
in the DAG.  Therefore, this may clearly
not be the correct choice as Example~\ref{example:motivating}
demonstrates.
}

\eat{Consider the following example.
Suppose the optimal plans for joining $R$, $S$ and $P$ is
$ (R \Join S) \Join P$  and the optimal plan for joining 
$R$, $S$ and $T$ is $ (R \Join T) \Join S$.
No common subexpression is present, and thus the Volcano-SH algorithm
will not introduce any sharing.
However, when optimizing the join of $R$, $S$ and $T$, if we realized that
$R \Join S$ is already present in an earlier plan and could be reused,
we may find that the plan $(R \Join S) \Join T$ is cheaper than
$ (R \Join T) \Join S$.
}

% If common sub--expressions are present in a plan, the sub--expression
% can be materialized and shared. The cost of evaluating the common
% sub--expression should thus be counted only once in the evaluation cost
% of the plan.  While computing the cost of the candidate plans for the
% (physical) equivalence nodes, basic Volcano however does not take into 
% account the above;
% and thus the best plan chosen is not necessarily the best plan globally.
% \subsubsection{Implementation of Volcano-RU}
\fullversion{
In the Volcano search algorithm, when finding the best plan for an 
equivalence node we recursively compute the best plans of each of its
children equivalence nodes.
Intuitively, our extended algorithm, which we call Volcano-RU, 
uses the same algorithm with one change:
it keeps track of what equivalence nodes are included in the current
partial plan being explored, and treats all such nodes as being 
materialized and available for reuse.
Thus the cost of using the node is the minimum of its reuse cost and
its recomputation cost.
}

%\subsubsection{The Volcano-RU algorithm}

The intuition behind Volcano-RU is as follows.
Consider $Q_1$ and $Q_2$ from Example~\ref{example:motivating}.
With the best plans as shown in the example, namely
$(R \Join S) \Join P$ and $(R \Join T) \Join S$,
no sharing is possible with Volcano-SH.
However, when optimizing $Q_2$, if we realize that $R \Join S$ is
already used in in the best plan for $Q_1$ and can be shared, 
the choice of plan $(R \Join S) \Join T$ may be found to be the best for
$Q_2$.

The intuition behind the Volcano-RU algorithm is therefore as follows.
Given a batch of queries, Volcano-RU optimizes them in
sequence, keeping track of what plans have already been chosen for 
earlier queries, and considering the possibility of reusing parts of 
the plans.  
The resultant plan depends on the ordering chosen for the queries; we 
return to this issue after discussing the Volcano-RU algorithm.

\begin{figure}
\begin{small}
\ordinalg{
Procedure {\sc Volcano-RU} \\
{\em Input:} \> \> Expanded DAG on queries $Q_1, \ldots, Q_k$ (including 
subsumption derivations) \\
{\em Output:} \> \> Set of nodes to materialize $M$,
			and the corresponding best plan $P$ \\
\> $N$ = $\phi$ // Set of potentially materialized nodes \\  
\> For each equivalence node $e$, Set $count[e]=0$ \\
\> For $i = 1$ to $k$ \\
\> \> Compute $P_i$, the best plan for $Q_i$, using Volcano, assuming 
% \\ \> \> \> 
      nodes in $N$ are materialized \\
\> \> For every equivalence node in $P_i$ \\
\> \> \> set $count[e] = count[e]+1$ \\
\> \> \> If ($cost(e) + matcost(e) + count[e]*reusecost(e) < 
                    (count[e]+1)*cost(e)$) \\
\> \> \> \> \> // Worth materializing if used once more \\
\> \> \> \> add $e$ to set $N$ \\
\> Combine $P_1, \ldots, P_k$ to get a single DAG-structured plan $P$ \\
\> (M,P) = {\sc Volcano-SH}($P$)  ~~~~~~// Volcano-SH makes final
              materialization decision
}
\end{small}
\vspace{-2mm}
\caption{The Volcano-RU Algorithm}
\label{fig:volcano-ru}
\end{figure}

The pseudocode for the Volcano-RU algorithm is shown in
Figure~\ref{fig:volcano-ru}.
Let $Q_1, \ldots, Q_n$ be the queries to be optimized together
(and thus under the same pseudo-root of the DAG).  
The Volcano-RU algorithm optimizes them in the sequence $Q_1, \ldots, Q_n$.
After optimizing $Q_i$, we note equivalence nodes in the 
DAG that are part of the best plan $P_i$ for $Q_i$ as candidates
for potential reuse later.  
We maintain counts of number of uses of these nodes.
We also check if each node is worth materializing, if it is used
one more time.
If so, we add the node to $N$, and when optimizing the next
query, we will assume it to be available materialized.

Thus, in our example earlier in this section, after finding the best
plan for the first query, we check if $R \Join S$ is worth materializing 
if it is used once more.  If so we add it to $N$, and assume it to
be materialized when optimizing the second query.

After optimizing all the individual queries, the second phase of Volcano-RU
executes Volcano-SH on the overall best plan found as above to further
detect and exploit common subexpressions.  This step is essential since
the earlier phase of Volcano-RU does not consider the possibility of
sharing common subexpressions within a single query -- equivalence nodes
are added to $N$ only after optimizing an entire query.
Adding a node to $N$ in our algorithm does not imply it will get reused and
therefore materialized.  Instead Volcano-SH makes the final decision
on what nodes to materialize.  The difference from directly applying
Volcano-SH to the result of Volcano optimization is that the plan $P$
that is given to Volcano-SH has been chosen taking sharing of parts of 
earlier queries into account, unlike the Volcano plan.

\reminder{See if we can use the earlier example for this as well?}

A related implementation issue is in caching of best plans in the DAG.
When optimizing $Q_i$ we cache best plans in nodes of the DAG that 
are descendants of $Q_i$.
When optimizing a later query $Q_j$, if we find a node that is not in
$P_i$ (the plan chosen for query $Q_i$) for some $i < j$, we must 
recompute the best plan for the node;
for, the set of nodes $M$ may have changed, leading to a different 
best plan.
Therefore we note with each cached best plan which query was being optimized
when the plan was computed; we recompute the plan as required above.

Note that the result of Volcano-RU depends on the order in which
queries are considered.  
In our implementation we consider the queries in the order in
which they are given, as well as in the reverse of that order,
and pick the cheaper one of the two resultant plans.
% As a simple extension, we can consider the queries in forward and reverse
% order and find the minimum cost plan amongst the two.
Note that the DAG is still constructed only once, so the extra cost of
considering the two orders is relatively quite small.
Considering further (possibly random) orderings is possible, 
but the optimization time would increase further.

%%%%%%%%%%%% END OF SECTION %%%%%%%%%%%%%%%%%%%%%

\eat{%% SUD: Stuff I ate now
%%%%%%%%%%%%%%%%%%%%%%%%%%%%%%%%%%%%%%%%%%%%%%%%%%%%%%%%%%%%%%%%%%%%%%%%%%%
The first stage of Volcano-RU is identical to the Volcano search
strategy, but with the following changes: 
\begin{enumerate}
\item Each node carries a flag stating whether it is to be materialized
or not.  When a marked node is encountered while optimizing query $Q_i$,
its cost is computed as follows. If the node has already been flagged as
materialized, then its cost is taken as the reuse cost of the materialized
result.  Otherwise its recomputation cost is compared with the sum of
its materalization cost and reuse cost (accounting for materialization
on the first computation and a reuse now) and the minimum taken as the
cost of the node; if the latter is the minimum then the node is flagged
as materialized as well.
\item If during the course of optimization of a query $Q_i$, the best plan
for a node is computed but is not a part of the overall best plan of
$Q_i$ (and, therefore, does not get marked), then this best plan is
discarded after
optimization of $Q_i$ is completed, and a new best plan is computed
\eat{
not memoized and the best plan is computed anew
}
if this node is encountered later during the
optimization of a query $Q_j, j>i$.
\eat{
Since optimization is carried out an a DAG with shared nodes,
we may revisit an equivalence node when optimizing different queries.
When we visit an equivalence node from $Q_{i}$, if the node had 
been optimized as part of the optimization process for $Q_j$, $j < i$,
but is unmarked (and thus not part of the best plan of any query so far) 
we must not reuse the old cached plan at that node, and must
reoptimize the equivalence node.
}
The reason is that more nodes may have been marked since 
$Q_i$ was optimized and they may provide a cheaper way of 
computing the equivalence node when $Q_j$ is being optimized.
\eat{
Therefore, we note with each equivalence node the query during
which its best plan was last computed; if it was not during the current
query being optimized, we discard the stored best plan and recompute it.
}
\end{enumerate}

The Volcano search algorithm, with the new definition of cost for 
marked nodes and the reoptimization of unmarked nodes, guarantees that the 
plan chosen for a query $Q_i$ is the best, assuming the (marked) nodes in 
the best plans of queries $Q_j, j <i$ can be shared, and no other nodes 
can be shared.
%%%%%%%%%%%%%%%%%%%%%%%%%%%%%%%%%%%%%%%%%%%%%%%%%%%%%%%%%%%%%%%%%%%%%%%%%%%
}

\eat{
Unlike Volcano-SH, Volcano-RU will pick plans that
use subsumption derivations across two different queries. 
Thus, subsumption-derivations should be present in the DAG used by 
Volcano-RU during the initial optimization phase, 
in addition to being considered again by the Volcano-SH post-pass.

The above description ignored the materialization cost of marked
nodes during the initial phase of Volcano-RU.
To take materialization cost into account, we reuse a marked node
from an earlier query only if its materialization cost plus its reuse cost 
is less than its recomputation cost.  At the end we add up the 
materialization costs for all reused nodes to the total cost.
}
% Note, that within a query,  Volcano-RU will
% not pick plans with subsumption derivations but Volcano-SH
% will take care of this in its post pass.

%\subsubsection{Discussion}

\eat{
We also attempted to extend Volcano-RU to consider reuse of common 
subexpressions {\em within} a single query.
However, this turned out to be quite difficult, since until the choice
of the best alternative at the top of the query DAG is made, we don't know 
which nodes in the DAG below are part of the best plan and which are not.
We developed extensions of the Volcano algorithm that consider reuse within
a query, but found them to be very expensive in time and space.  The high
cost appears to be intrinsic to the problem, hence we dropped this extension
from consideration.
}

\fullversion{
However, there are several efficiency problems due to the fact that 
Volcano explores the
search space by considering various alternatives and discarding 
alternatives that are not better than the best plan (for each
equivalence  node) known so far.
Thus, equivalence nodes are considered and dropped from the optimal 
plan (so far) as optimization proceeds. 
We developed a marking procedure that keeps track of the current
plan being considered by Volcano, marking and unmarking equivalence 
nodes as appropriate.
However, as outlined
earlier, if the set of marked nodes when the best plan for an equivalence
node was found is different from the current set, the best plan cannot be 
reused. Therefore, one choice is to reoptimize afresh each time
we visit an equivalence node. However, this is prohibitively expensive
since it throws away dynamic programming.
\fullversion{
For example, if we are optimizing, $R1 \Join R2 \ldots Rn$,
the equivalence node corresponding to $R1 \Join R2 \Join R3$
may be optimized $2^{n-3}$ times, once for each subexpression
it belongs to.
}
The other choice is to use a caching scheme that keeps track 
of multiple best-plans for each equivalence node, corresponding to different
sets of marked nodes.  
We developed and implemented these extensions  but found that they 
performed quite badly in terms of time, and the space needed for caching
was excessive.
}

\eat{ %%% I think this does not contribute much %%%
\subsections{Reuse Based Exhaustive Algorithms}

Although Volcano-RU makes reuse decisions based on what has been used 
earlier, it is not able to predict what will be used later in the query.  
For example, if we were to swap $Q_1$ and $Q_2$ in 
Example~\ref{example:motivating}, it is easy to see that 
Volcano-RU will fail to come up with the globally optimal plan.
In fact, Volcano-RU will not obtain the globally
optimal plan in Example~\ref{example:motivating}, 
{\em irrespective} of the order in which the queries are
optimized, if the local best plan for $Q_1$ is 
$(R \Join P) \Join S$ and the globally optimal
plan involves using $R \Join S$.
%Further, just like Volcano-SH, its detection of common subexpressions
%is only done on the best plan produced by the first phase;
%the first phase, however, chooses plans in ignorance of 
%potential benefits of common subexpression sharing within a query.
%%%%%%%%%%%%%%
% It may therefore make the wrong choice
% of a plan initially, which prevents sharing with later parts of the query.
% We also outline an algorithm to extend Volcano-RU to 
% consider reuse of common subexpressions {\em within} a single query
% in~\cite{rssb:tr}.  These algorithms are however very expensive.
% In fact, the high cost appears to be intrinsic to the problem
% as discussed in~\cite{rssb:tr}. 
%%%%%%%%%%%%%%
In~\cite{rssb:tr},
%In Appendix~\ref{app:exh:backtrack},
we outline an optimal exhaustive algorithm, based
on backtracking, which considers plans that are not locally optimal,
and accounts for reuse of nodes in the current plan.
Although our implementation of this algorithm worked well on small
queries, like all exhaustive algorithms for this problem, it is
not practical for large queries.

% as demonstrated by our experiments in Section~\ref{sec:perf}.

\eat{
For a global optimum, we would have to take into consideration that 
a plan may allow sharing with later parts of the overall query, even 
if it is more expensive locally.   
In terms of our DAG structure, a particular (physical) operation node may
be best for implementing a (physical) equivalence node, although
it may not be the cheapest, since one of its descendant equivalence nodes
can be shared with an earlier or a later part of the query.
Thus, for each equivalence node, we cannot choose just the cheapest
operation node, but must explore non-cheapest alternatives as well.

We have developed the above idea into an exhaustive search algorithm
based on backtracking.  
\fullversion{
Readers familiar with the Prolog search strategy
will note some similarities:  each rule corresponds to an AND-node, while 
each predicate has a set of alternative rules defining it, and thus 
corresponds to an OR-nodes.  The analogy must not be stretched too far,
since there are differences.  However using the Prolog backtracking 
search algorithm as an analogy, we have developed our backtracking search
algorithm.  
}
We have also introduced a number of optimizations into our exhaustive
backtracking algorithm.  One optimization detects if no descendant of 
an equivalence node is sharable, and if so it simply chooses the cheapest
operation node child for it, without backtracking.  
\fullversion{
The optimization is
refined further to allow some child operation nodes to have sharable
descendants, and restrict backtracking to these nodes.
}
Further, caching of best plans is done so that if a search state is revisited
by the backtracking algorithm the earlier best plan can be reused in 
some cases, where the set of sharable nodes from earlier in the plan is
the same as in an earlier visit.  

\fullversion{
Space does not permit us to give full details of the exhaustive
backtracking algorithm.   Like any exhaustive algorithm for this problem, 
it can become very expensive as the size of the query to be optimized 
increases. It is not practical for medium to large queries, and 
therefore not a serious contender for multi-query optimization in the 
real world.
}

Like any exhaustive algorithm for this problem, the exhaustive 
backtracking algorithm is not practical for large queries.
However, due to the above-mentioned optimizations, the algorithm
is often very efficient when the amount of sharing is low.   In fact,
we used it to generate the ``optimal cost'' figures for
many of our the experiments in our performance study.
}
}

%%%%%%%%%%%%%%%%%%%%%%%%%%%%%%%%%%%%%%%%%%%%%%%%%%%%%%%%%%%%%%%%%%%%%%%%%%%%

%% file: greedy.tex
\sections{The Greedy Algorithm}
\label{sec:greedyalgo} 

In this section, we present the greedy algorithm, which provides
an alternative approach to the algorithms of the previous section.
Our major contribution here lies in how to 
{\em efficiently implement} the greedy algorithm, and we shall concentrate
on this aspect.

In this section, we present an algorithm with a different optimization
philosophy. The algorithm picks a set of nodes $S$ to be materialized
and then finds the optimal plan given that nodes in $S$ are materialized.
This is then repeated on different sets of
nodes $S$ to find the best (or a good) set of nodes to be materialized.

% In this section we assume a single query $Q$ as the input instead of
% the set of queries $Q_1, \ldots, Q_k$ as considered in the previous
% section. This is essentially for ease of presentation; 

Before coming to the greedy algorithm, we present some definitions,
and an exhaustive algorithm.
As before, we shall assume there is a virtual root node for the DAG;
this node has as input a ``no-op'' logical operator whose inputs are
the queries $Q_1 \ldots Q_k$. 
Let $Q$ denote this virtual root node.

For a set of nodes  $S$, let $bestcost(Q, S)$ denote the cost of the
optimal plan for $Q$ given that nodes in $S$ are to be materialized
(this cost includes the cost of computing and materializing nodes in $S$).
As described in the Volcano-SH algorithm, the basic Volcano optimization
algorithm with an appropriate definition of the cost for nodes in $S$
can be used to find $bestcost(Q, S)$.  

To motivate our greedy heuristic, we first describe a simple exhaustive 
algorithm.
The exhaustive algorithm, iterates over each subset $S$ of the set of 
nodes in the DAG, and chooses the subset $S_{opt}$ with the minimum 
value for $bestcost(Q, S)$. 
Therefore, $bestcost(Q, S_{opt})$ is the cost of the globally
optimal plan for $Q$.

It is easy to see that the exhaustive algorithm is doubly
exponential in the size of the initial query DAG and is
therefore impractical. 

In Figure~\ref{fig:greedy} we outline a greedy heuristic
that attempts to approximate $S_{opt}$ by constructing 
it one node at a time.  The algorithm iteratively picks nodes
to materialize.  At each iteration, the node $x$  that gives the 
maximum reduction in the cost if it is materialized is chosen 
to be added to $X$. 

\begin{figure}
\begin{small}
\ordinalg{
Procedure {\sc Greedy} \\
{\em Input:} \> \> Expanded DAG for the consolidated input query $Q$ \\
{\em Output:} \> \> Set of nodes to materialize
and the corresponding best plan \\
\> X = $\phi$ \\
\> Y = set of equivalence nodes in the DAG \\
\> while (Y $\neq \phi$) \\
L1:\>   \> Pick the node x $\in$ Y with the smallest value for  bestcost(Q, \{x\} $\cup$ X) \\
\> \>  if (bestcost(Q, \{x\} $\cup$ X) $<$ bestcost(Q, X) ) \\
\> \>    \>Y  = Y - x;~~~X = X $\cup$ \{x\} \\
\> \>  else Y = $\phi$ // benefit $<$ 0, so break out of loop \\
\> return X
}
\end{small}
\vspace{-2mm}
\caption{The Greedy Algorithm}
\label{fig:greedy}
\end{figure}

The greedy algorithm as
described above can be very expensive due to the large number of
nodes in the set $Y$ and the large number of times the function
$bestcost$ is called. 
We now present three important and novel optimizations
to the greedy algorithm which make it efficient and practical.
%as demonstrated by our experiments in Section~\ref{sec:perf}.
\begin{enumerate}
% \reminder{actually we need to formally define a plan ...}
\item The first optimization is based on the observation that
the nodes materialized in the globally optimal plan
are obviously a subset of the ones that are shared in some
plan for the query. Therefore, it is sufficient to 
initialize $Y$ in Figure~\ref{fig:greedy},
with  nodes that are shared in some plan for the query.
We call such nodes {\em sharable nodes}.
For instance, in the expanded DAG for $Q_1$ and $Q_2$ corresponding to 
Example~\ref{example:motivating}, $R \Join S$ is sharable
while $R \Join T$ is not.
We present an efficient algorithm for finding  
sharable nodes in Section~\ref{sec:sha}.
%Appendix~\ref{app:sha}.
%~\cite{rssb:tr} but
%do not present the details here for lack of space. 

%In our experiments in Section~\ref{sec:perf},
%we observed that this resulted in a dramatic reduction in
%the number of nodes that $Y$ is inititialized with and
%consequently a dramatic reduction in the cost of the
%greedy algorithm itself.

\eat{
At each step the algorithm chooses the node $x \in Y$ with the smallest
value for $bestcost(Q, \{x\} \cup X)$,
where $bestcost(Q, S)$ is the cost of the optimal plan given that
nodes in $S$ have been chosen to be materialized. We outline how
to compute $bestcost$ in Section~\ref{ssec:mat:queryopt}.
}

\item The second optimization is based on the
observation that there are many calls to 
$bestcost$ at line L1 of Figure~\ref{fig:greedy},
with different parameters.
A simple option is to process each call to $bestcost$ independent 
of other calls. 
However, observe that the symmetric difference\footnote{The symmetric
difference of two sets $S_1$ and $S_2$ consists of elements that 
are in one of the two but not both; formally the symmetric difference
of sets $S_1$ and $S_2$ is $(S_1 - S_2) \cup (S_2 - S_1)$,
where $-$ denotes set difference.
}
in the sets passed as parameters to successive calls to 
$bestcost$ is very small -- sucessive calls take
parameters of the form $bestcost(Q, \{x\} \cup X)$, where only $x$ varies.
It makes sense for a call to leverage the work done by a previous call. 
We describe a novel incremental cost update algorithm,
in Section~\ref{sec:inc}, that maintains the state of the optimization 
across calls  to $bestcost$, and 
incrementally computes a new state from the old state.

\item The third optimization, which we call the monotonicity heuristic,
avoids having to invoke $bestcost(Q,$ $\{x\} \cup X)$, for 
every $x \in Y$, in line L1 of Figure~\ref{fig:greedy}.
We describe this optimization in detail in Section~\ref{sec:monotonicity}.

\end{enumerate}
\eat{
Clearly, one could compute bestcost(Q, S), for
every subset of $Y$ in Figure~\ref{fig:greedy}
and thus find the optimal plan by an exhaustive search.
However, the exhaustive search is extremely 
expensive since the set of sharable nodes itself may be exponential in the 
size of the initial query, and the set of all subsets of this set
may be doubly exponential in the size of the query. 
Other variants of exhaustive searching, such as A$^*$ search
as proposed in \cite{tim:mul,lqa97:phys} are also possible,
but suffer from the same problems.
It is also worth noting that one could also choose any other heuristic to 
decide which subsets of $Y$ to consider for materialization. These
heuristics can also benefit from the above mentioned optimizations. 
}

\eat{
\subsection{Finding Optimal Plans With Materialized Expressions} 
\label{ssec:mat:queryopt}

Suppose a set $S$ of (physical) equivalence nodes have been
selected to be materialized.
We now address the task of finding the optimal plan given this decision,
i.e., the task of finding $bestcost(Q, S)$.
This task has two parts: (a) finding the best overall query plan
given the results of nodes in $S$ have been materialized,
(b) finding the best plan for computing each node in $S$, assuming the 
other nodes in $S$ are materialized.
The overall query cost given all nodes in $S$ are materialized,
$bestcost(Q, S)$ is then the cost of the plan in part (a) plus the cost computed
in part (b) above for every $s$ in $S$.

The solution to part (a) above is similar to the
algorithm used in Volcano-RU.  We define the cost of a equivalence node
in the set $S$ to be the minimum of the reuse and recomputation cost
and use the basic Volcano search strategy along with this
new definition of the cost of a node in $S$.

% ({\em For the benefit of referees, the above procedure is outlined in
% Appendix~\ref{app:bestplan}.})

% For part (b),  when computing the cost of a node $s$ in $S$,
% repeat the procedure outlined as above assuming that all nodes in
% $S$ other than $s$ have been materialized.
%%%%%%%%%%%
% The solution to part (b) above is correct if
% there are no cyclic derivations between nodes in $S$, i.e.,
% if a node $a$ in $S$ can be used to compute $b$ in $S$, the
% converse is not possible. 
% If the nodes in $S$ involved logical eauivalquivalence nodes
% only, then such cyclic derivations are not possible in the
% logical DAG.

The solution for part (b) is complicated by the fact that 
cyclic derivations are possible in the physical DAG.
For instance, $r \Join s$ sorted on attribute $A$ and $B$ can be derived
from $r \Join s$ sorted on attribute $A$ and $C$, as well as vice versa,
by means of sort enforcers.
If we choose both the physical expressions to be materialized,
and are careless in choosing the overall plan, 
the best plan for one may be found to be obtained by sorting the result
of the other, and vice-versa.  Clearly such a cyclic plan cannot be
implemented.

Luckily such cycles are only possible between physical equivalence
nodes corresponding to a single logical equivalence node.
Our optimization algorithms take care of this possibility,
by means of executing a minimum spanning tree based algorithm
that avoids creating plans with cycles.
We omit the details here to keep the paper concise, and details may be
found in~\cite{rssb:tr}.

\fullversion{
({\em For the benefit of the referees, we outline the ideas behind
the algorithm in Appendix~\ref{app:phys:subsum}.})
}

\fullversion{
At the logical expression level we shall assume that no such cycles
are possible.
Indeed normal transformation rule systems will ensure this property.
There are some unusual ways of deriving expressions where this does not
hold:  for instance, consider $r \Join s$ and $ r \Join s \Join t$,
where the $\Join$ is a natural join.
Obviously the latter can be computed from the former; 
the former can be computed from the latter in the special case that
the join attribute of $t$ is a foreign key from $r \Join s$.
We rule out such transformations for simplicity of exposition.
}

It is worth nothing that part (a) of our algorithm above
is actually a solution for the general problem of query optimization 
in the presence of previously materialized views.
%Prior solutions we are aware of, such as~\cite{surajit:matviewopt},
%only address select/project/join queries, whereas our solution
%can handle a more general class of queries, including 
%aggregation and nested queries.\footnote{A full implementation would include 
%more transformation rules for groupby, including heuristics to 
%limit the search space~\cite{surajit:groupbyopt,larson:eageragg}.
%These have been
%incorporated into our basic optimizer \cite{sri:agg}, but 
%have not yet integrated with multi-query optimization.
%}
%Sesh: ate the above -- not sure that surajit does
% not consider group by -- in fact I think he does.
% also messy to discuss what we have and have not
% implemented in the concepts section -- anyway related 
% work has taken care of this
\eat{
The final evaluation plan is obtained by removing
redundant expressions in $S$, namely expressions which do not 
form a part of the consolidated optimal plan of the root).
}
}

\subsections{Sharability}
\label{sec:sha}

\eat{
We define a logical equivalence node to be {\em shared} in a plan if there are
two different paths from the root of the plan to that node.
We say that a node is {\em sharable} in an expanded logical AND-OR DAG if there 
exists some plan for the input query in which the node is shared.
}

In this subsection, we outline how to detect whether an equivalence
node can be shared in some plan.
The plan tree of a plan is the tree obtained from the DAG structured plan,
by replicating all shared nodes of the plan, to completely eliminate sharing.
The degree of sharing of a logical equivalence node in an evaluation 
plan $P$ is the number of times it occurs in the plan tree of $P$. 
The {\em degree of sharing} of a logical 
equivalence node in an expanded DAG
is the maximum of the degree of sharing of the equivalence node amongst
all evaluation plans represented by the DAG.
A logical equivalence node is {\em sharable} if 
its degree of sharing in the expanded DAG is greater than one.
\eat{
The degree of sharing and sharability of a physical equivalence node
are inherited from its associated logical equivalence node.%
\footnote{It is possible to directly compute the degree of sharing
of physical equivalence nodes, and we had done so in an earlier implementation.
However, since the physical DAG is significantly bigger than the logical
DAG, performing the computation for logical equivalence nodes and using 
the result for corresponding physical equivalence nodes is much
cheaper.}
}

%Figure ~\ref{fig:deg} presents an algorithm for computing the 
%degree of sharing of an equivalence node in a DAG.
We now present a simple algorithm to compute the degree
of sharing of each node and thereby detect whether a node is shared. 
A sub--DAG of a node $x$ consists of the nodes below 
\eat{reachable from} $x$ along with the edges between these
nodes that are in the original DAG.
For each node $x$ of the DAG, and every node $z$ in the sub-DAG rooted
at $x$, let $E[x][z]$ represent the degree of sharing of $z$ in
the sub--DAG rooted at $x$.
Clearly for all nodes $x$, $E[x][x]$ is $1$.
For a given node $x$, all other $E[x][\_]$ values can be computed 
given the values $E[y][\_]$ for all children $y$ of $x$ as follows.

If $x$ is an operation node \\
\hspace*{1cm}$ E[x][z] = Sum\{E[y][z] ~|~ y \in children(x)\}  $ \\
and if $x$ is an equivalence node, \\
\hspace*{1cm}$  E[x][z] = Max\{E[y][z] ~|~ y \in children(x)\}  $ \\
%%%%%%%%%%%
% If $x$ is an operation node, $E[x][z]$ is obtained by adding $E[y][z]$
% for all children $y$ of $x$.
% If $x$ is an equivalence node, $E[x][z]$ is obtained by taking the
% the maximum of $E[y][z]$, for all children $y$ of $x$.
%%%%%%%%%%%
The degree of sharing of a node $z$ in the overall DAG is given by
$E[r][z]$, where $r$ is the root of the DAG.

Space is minimized in the above by computing $E[x][z]$ for one $z$
at a time, discarding all but $E[r][z]$ at the end of computation for
one $z$ value.
\eat{ %% very conservative estimate
The above procedure has a worst case time complexity
of $O(ne)$ where $n$ is the number of nodes in the DAG, and $e$ is the
number of edges in the DAG.  However, this reduces overall cost, as
illustrated by our performance experiments in Section~\ref{sec:perf},
since it reduces the number of nodes considered for materialization in
Figure~\ref{fig:greedy}.
}

A reasonable implementation of the above algorithm has time complexity
proportional to the number of non-zero entries in $E$, which in the worst
case is proportional to the square of the number of nodes in the DAG.  
However, typically, $E$ is fairly sparse since the DAG is typically
``short and fat'' -- as the number of queries grows, the height of the
DAG
%% does \reminder{strong statement}
may not increase, but it becomes wider.
Thus, as expected, this sharability computation algorithm is fairly efficient
in practice.  In fact, for the queries we considered in our performance
study (Section~\ref{sec:perf}), the computation took at most a few tens
of milliseconds.

% Note that the worst case complexity of the greedy algorithm
% is $O(k^2p)$ where $p$ is the complexity of finding
% $bestcost$ and as we will see shortly is in the worst case
% $O(e)$. Thus, reducing $k$, is worthwhile; 
% in practise we have observed that the benefits of this procedure have 
% proven to be worth its cost.
% \reminder{See if we can substantiate with numbers}
% \reminder{need to check above analysis}

\eat{
\begin{figure}[thbp]
\renewcommand{\baselinestretch}{1.05}
\ordinalg{
Inputs: \> \> children(x) : children of a node x\\
\> \> R : set of base relations\\
Output: \> \> E[x][y] = degree of sharing of node y in a DAG rooted at x\\
\\
Function FindDegree\\
for each node x in R\\
\> for each node z \\
\> \> if (z = x) \\
\> \> \> E[x][z] = 1\\
\>\>  else E[x][z] = 0\\
S = R \\
while there exists a node x not in S such that children(x) is a subset of S \\
\> for each node z \\
\> \> if (z = x) \\
\> \> \> E[x][z] = 1  \\
\> \> else if x is an AND node \\
\>\>\>      E[x][z] = Sum\{E[y][z] $|$ y in children(x)\}  \\
\> \> else /* x is an OR node */\\
\>\>\>      E[x][z] = Max\{E[y][z] $|$ y in children(x)\}\\
\> S = S $\cup$ \{x\} 
}
\vspace{-0.1in}
\caption{Computation of the Degree of Sharing}
\label{fig:deg}
\end{figure}
}

\subsections{Incremental Cost Update}
\label{sec:inc}

The sets with which $bestcost$ is called successively at
line L1 of Figure~\ref{fig:greedy} are closely related, 
with their (symmetric) difference being very small.
For, line L1 finds the node $x$ with the maximum benefit, which
is implemented by calling $bestcost(Q, \{x\} \cup X)$, for different
values of $x$.
Thus the second parameter to $bestcost$ changes by dropping one node $x_i$
and adding another $x_{i+1}$.
\eat{
We now present an incremental cost update algorithm that 
exploits the results of earlier cost computations to incrementally 
recompute the optimal plan.
}
We now present an incremental cost update algorithm that exploits the
results of earlier cost computations to incrementally compute the new
plan. 
\eat{
%%%%%%% Sud: commented out .... %%%%%%%
The new plan will be optimal if we disallow pipelining the first
computation of a shared result -- this is because the cost of using the
result is the same for each parent. If we allow pipelining, then the
first use of the shared result is at the cost of computing the result
and simultaneously materializing it, while the remaining uses are at the
cost of using the materialized result; since the order of computation is
not known apriori, the new plan generated incrementally is a heuristic
approximation. Results from the performance study in Section~\ref{sec:perf}
show that this heuristic works well for the queries we considered.
%%%%%%%%%%%%%%%%%%%%%%%%%%%%%%%%%%%%%%%%%
}
 
\begin{figure}% [thbp]
{\small 
\ordinalg{
\renewcommand{\baselinestretch}{1.10}
Function UpdateCost($S$, $S'$)\+\\
// PropHeap is a priority heap (initially empty), containing
// equivalence nodes ordered by their topological sort number
add $S-S' \union S'-S$ to PropHeap\\
while (PropHeap is not empty) \\
\> $N$ = equivalence node with minimum topological sort number in PropHeap \\
\> Remove $N$ from PropHeap \\
\> oldCost = old value of cost($N$) \\
\> cost($N$) = Min $\{$ cost($p$) | $p \in  children(N) \} $
	// $children(N)$ are operation nodes \\
\> if (cost($N$) $\neq$ oldCost) or $N \in (S - S')$ or $N \in (S' - S)$ \\
\> \> for every parent operation node $p$ of $N$ \\
\> \> \> cost($p$) = cost of executing operation $p$ + 
	$\sum_{c \in children(p)}(C(c))$ \\
\> \> \> \> where $C(c)$ = cost($c$) if $c \not\in S'$, and 
	$ = min(reusecost(c),$ cost($c$)) if $c \in S'$ \\
\> \> \> add $p$'s parent equivalence node to PropHeap if not already present \\
TotalCost = $compcost(root) + \sum_{s \in S'}($cost($s$) $+ matcost(s))$ \-
}
}
\caption{Incremental Cost Update}
\label{fig:inc}
\end{figure}

Figure~\ref{fig:inc} outlines our incremental cost
update algorithm. 
Let $S$ be the set of nodes shared at 
a given point of time, i.e., the previous call to $bestcost$
was with $S$ as the parameter. The incremental cost
update algorithm maintains the cost of 
computing every equivalence node, given that all nodes in $S$ are shared,
and no other node is shared. Let $S'$ be the new set of nodes that 
are shared, i.e., the next call to $bestcost$ has $S'$
as the parameter. 
The incremental cost
update algorithm starts from the nodes that have changed in going 
from $S$ to $S'$ (i.e., nodes in $S'-S$ and $S-S'$) and
propagates the change in cost for the nodes upwards
to all their parents; these in turn propagate any changes in cost to 
their parents if their cost changed, and so on, until there is no 
change in cost. 
Finally, to get the total cost we add the cost of computing and materializing
all the nodes in $S'$.

\reminder{May want to give an example here....}

\eat{
At each node, the cost is recomputed in the same fashion as in the
non-incremental algorithm described in the previous section.\footnote{
All physical equivalence nodes corresponding to a logical
equivalence node are processed together, when the logical equivalence
node is popped from the heap.  The minimum-spanning-tree based 
algorithm is used to process cyclic derivations between the nodes, 
as mentioned earlier.
}
}

If we perform this propagation in an arbitrary order then in the worst case we
could propagate the change in cost through a node $x$ multiple times (for
example, once from a node $y$ which is an ancestor of another node $z$
and then from $z$). 
\fullversion{
So, at any point of time we have a set $P$ of nodes
from whom a cost change has to be propagated (initially it is $(S'-S)
\union (S-S')$). We choose the next node from which cost is propagated
as one for which no descendant of it is in $P$.  
}
A simple mechanism for avoiding repeated propagation uses 
topological numbers for nodes of the DAG.
During DAG generation the \fullversion{logical} DAG 
is sorted topologically such that a
descendant always comes before an ancestor in the sort order, and
nodes are numbered in this order. 
As shown in Figure~\ref{fig:inc},
cost propagation is performed in the topological number ordering
using {\em PropHeap}, a heap built on the topological number.
The heap is used to efficiently find the node with the minimum topological 
sort number at each step.

In our implementation, we additionally take care of physical property
subsumption.  Details of how to perform incremental cost update
on physical DAGs with physical property subsumption are given in
\cite{rssb:tr}.

\fullversion{
{\em For the benefit of the referees, the above details are shown in
Appendix~\ref{app:phys:subsum}.}
}

\fullversion{
Initially (before the greedy or the exhaustive algorithm 
starts execution) $S$ is empty. 
Therefore, initially the cost of computing a operation node given that
nothing is materialized is computed using the basic Volcano algorithm. 
}

\eat{
\subsection{Exhaustive Algorithm Based on Sharable Nodes}
\label{sec:exh:sha}

Based on the observations in the preceding sections, we can find 
a globally optimal plan by simply doing the following:
for every subset $S$ of the set of sharable nodes, compute $bestcost(Q, S)$.
Choose the set $S_{min}$ that gives the minimum plan cost;
the plan chosen for $S_{min}$ constitutes a globally optimal plan 
for the entire query.   

Rather than explore the subsets $S$ in an arbitrary order, we can
improve efficiency as follows.
The incremental cost update is most efficient when the
symmetric difference of $S$ and and a new set $S'$ is small, since
few costs will change in such a situation.
To find the best plan overall efficiently, therefore, we enumerate 
the subsets of the set of sharable nodes in {\em gray code order}.
Thereby, the symmetric difference between every pair of successive sets 
in the order is kept to the minimum, namely $1$.

In spite of the optimizations, the exhaustive algorithm 
is extremely expensive.  The set of sharable nodes itself 
may be exponential in the size of the initial query, and
the set of all subsets of this set may thus be doubly 
exponential in the size of the query! 
Other variants of exhaustive searching, such as A$^*$ search
as proposed in \cite{tim:mul,lqa97:phys} are also possible,
but suffer from the same problems.

\eat{
We can incorporate optimizations to this  very 
similar to those proposed in Section~\ref{sec:exh}  for effective 
cost-based pruning and prevention of re-computation of optimal 
subplans. However, we will not describe this for lack of space. 
Further, we do not consider this algorithm in the performance section and
consider only the exhaustive algorithm of Section~\ref{sec:exh}. 
}
}

\eat{
\subsection{Sharability}
\label{sec:sha}

% In this section, we will define the notion of a sharable node and
% degrees of sharing, on the expanded query DAG.
% We then present an algorithm for finding sharable nodes and maximum degrees 
% of sharing. 
% Some of our optimization algorithms presented later
% make use of the knowledge of which nodes are sharable and their 
% degrees of sharing
%%%%%%%%%%%%
% The intuition behind defining nodes as shared is as follows.
% Consider a plan; if a node in it has more than one parent, 
% convert the plan into a tree by repeatedly replicating nodes
% with multiple parents, and placing a separate copy under each parent.
% At the end, if a node appears more than once in the plan, it
% is shared in the plan.  A node is sharable if it is shared in any
% plan.
% 
%%%%%%%%%
% If a node is shared in any plan, its degree of sharing will be greater
% than 1 at the lowest operation node below which it appears twice or more,
% and at all ancestors of that operation node.
%%%%%%%%%

\eat{
We define a logical equivalence node to be {\em shared} in a plan if there are
two different paths from the root of the plan to that node.
We say that a node is {\em sharable} in an expanded logical AND-OR DAG if there 
exists some plan for the input query in which the node is shared.
}

We now outline how to detect whether an equivalence
node can be shared.
The plan tree of a plan is the tree obtained by replicating all 
shared subplans of the plan, to completely eliminate sharing.
The degree of sharing of a logical equivalence node in an evaluation 
plan $P$ is the number of times it can occur in the plan tree of $P$. 
The {\em degree of sharing} of a logical 
equivalence node in an expanded DAG
is the maximum of the degree of sharing of the equivalence node amongst
all evaluation plans represented by the DAG.
A logical equivalence node is {\em sharable} if 
its degree of sharing in the expanded DAG is greater than one.
The degree of sharing and sharability of a physical equivalence node
are inherited from its associated logical equivalence node.

\fullversion{
\footnote{It is possible to directly compute the degree of sharing
of physical equivalence nodes, and we had done so in an earlier implementation.
However, since the physical DAG is significantly bigger than the logical
DAG, performing the computation for logical equivalence nodes and using 
the result for corresponding physical equivalence nodes is much cheaper.
}
}

% Figure ~\ref{fig:deg} presents an algorithm for computing the 
% degree of sharing of an equivalence node in a DAG.

We now present a simple algorithm to compute the degree
of sharing of each node and thereby detect whether a node is shared. 
A sub--DAG of a node $x$ consists of the nodes below 
\eat{reachable from} $x$ along with the edges between these
nodes that are in the original DAG.
For each node $x$ of the DAG, and every node $z$ in the sub-DAG rooted
at $x$, let $E[x][z]$ represent the degree of sharing of $z$ in
the sub--DAG rooted at $x$.
Clearly for all nodes $x$, $E[x][x]$ is $1$.
For a given node $x$ $E(x,\_)$ values can be computed given the values 
$E(y,\_)$ for all children $y$ of $x$ as follows.
If $x$ is an operation node 
\[ E[x][z] = Sum\{E[y][z] $|$ y \in children(x)\}   \]
and if $x$ is an equivalence node,
\[  E[x][z] = Max\{E[y][z] $|$ y \in children(x)\}  \]
%%%%%%%%%%%
% If $x$ is an operation node, $E[x][z]$ is obtained by adding $E[y][z]$
% for all children $y$ of $x$.
% If $x$ is an equivalence node, $E[x][z]$ is obtained by taking the
% the maximum of $E[y][z]$, for all children $y$ of $x$.
%%%%%%%%%%%
The degree of sharing of a node $z$ in the overall DAG is given by
$E[r][z]$, where $r$ is the root of the DAG.

Space is minimized in the above by computing $E[x][z]$ for one
$z$ at a time, discarding all but $E[r][z]$ at the end of computation
for one $z$ value.
The above procedure has a worst case time complexity of $O(ne)$
where $n$ is the number of nodes in the DAG, and $e$ is the
number of edges in the DAG.  However, this reduces overall cost,
as illustrated by our performance experiments, since it
reduces the number of nodes considered for materialization 
in Figure~\ref{fig:greedy}.

% Note that the worst case complexity of the greedy algorithm
% is $O(k^2p)$ where $p$ is the complexity of finding
% $bestcost$ and as we will see shortly is in the worst case
% $O(e)$. Thus, reducing $k$, is worthwhile; 
% in practise we have observed that the benefits of this procedure have 
% proven to be worth its cost.
% \reminder{See if we can substantiate with numbers}
% \reminder{need to check above analysis}

\eat{
\begin{figure}[thbp]
\renewcommand{\baselinestretch}{1.05}
\ordinalg{
Inputs: \> \> children(x) : children of a node x\\
\> \> R : set of base relations\\
Output: \> \> E[x][y] = degree of sharing of node y in a DAG rooted at x\\
\\
Function FindDegree\\
for each node x in R\\
\> for each node z \\
\> \> if (z = x) \\
\> \> \> E[x][z] = 1\\
\>\>  else E[x][z] = 0\\
S = R \\
while there exists a node x not in S such that children(x) is a subset of S \\
\> for each node z \\
\> \> if (z = x) \\
\> \> \> E[x][z] = 1  \\
\> \> else if x is an AND node \\
\>\>\>      E[x][z] = Sum\{E[y][z] $|$ y in children(x)\}  \\
\> \> else /* x is an OR node */\\
\>\>\>      E[x][z] = Max\{E[y][z] $|$ y in children(x)\}\\
\> S = S $\cup$ \{x\} 
}
\vspace{-0.1in}
\caption{Computation of the Degree of Sharing}
\label{fig:deg}
\end{figure}
}

%\begin{figure}
%\centerline{\psfig{file=degree.ps}}
%\caption{Finding Degree of Sharing}
%\label{fig:deg}
%\end{figure}
}

\subsections{The Monotonicity Heuristic}
\label{sec:monotonicity}

In Figure~\ref{fig:greedy}, the function $bestcost$ will
be called once for each node in $Y$, under normal 
circumstances.  We now outline   
how to determine the node with the
smallest value of $bestcost$ much more efficiently, using the
monotonicity heuristic. 

Let us define
$benefit(x,X)$ as $bestcost(Q, X) - bestcost(Q, \{x\} \cup X)$.
Notice that, minimizing $bestcost$ in line $L1$
corresponds to maximizing benefit as defined here. 
Suppose the benefit is {\em monotonic}. 
Intuitively, the benefit of a node is monotonic if it never increases 
as more nodes get materialized; more formally
$benefit$ is monotonic if $\forall X \supseteq Y$,
$benefit(x,X) \leq benefit(x,Y)$.

\fullversion{
\reminder{Prasan: future work for you, can you put bounds on increases
of benefits in an anti-monotonic fashion, and make our heuristic impl.
always choose the most beneficial plan, even if monotonicity does not hold?
}
}

\fullversion{
This improvement is based on the observation that given the 
monotonicity property of the benefits, the benefit once computed for 
a given equivalence node always remains an upper bound on the 
actual benefit of the equivalence node. As such, as long as there
exists another equivalence node with actual benefit more than its previously
computed benefit, this node can never be the node with the greatest actual
benefit. Therefore, we need not recompute the benefit of a node unless
its previously computed benefit is the maximum among all equivalence nodes.
Further, if even after updation, the benefit of the node remains the
maximum among all nodes, it is clearly the maximum overall and the
equivalence node is chosen for materialization.
}

We associate an upper bound on the benefit of a node in $Y$ and maintain a
heap ${\cal C}$ of nodes ordered on these upper bounds.\footnote{This cost
heap is not to be confused with the heap on topological numbering used
earlier.} 
The initial upper bound on the benefit of a  node in $Y$
uses the notion of the maximum degree of sharing of the node 
(which we described earlier).
The initial upper bound is then just the cost of evaluating 
the node (without any materializations) times the maximum degree of sharing.
The heap ${\cal C}$  is now used to efficiently find the node 
$x \in Y$ with the maximum $benefit(x, X)$ as follows:
Iteratively, the node $n$ at the top ${\cal C}$ is chosen, its current
benefit is recomputed, and the heap ${\cal C}$ is reordered. 
If $n$ remains at the top, it is deleted from the ${\cal C}$ heap
and chosen to be materialized and added to $X$. 
Assuming the monotonicity property holds, the other values in the 
heap are upper bounds, and therefore, the node $n$ added to 
$X$ above, is indeed the node with the maximum real benefit.

If the monotonicity property does not hold, the node with maximum current
benefit may not be at the top of the heap ${\cal C}$ , but we still use 
the above procedure as a heuristic for finding the node with the greatest 
benefit.
Our experiments in Section~\ref{sec:perf} demonstrate that 
the above procedure greatly speeds up the greedy
algorithm.   Further, for all queries we experimented with,
the results were exactly the same even if the monotonicity heuristic
was not used.

%The monotonicity assumption is also made in, e.g., \cite{gupta97:viewsel}.
%As a heuristic, we use the procedure even if monotonicity may not hold.  
%This is repeated till the heap becomes empty or 
%the benefit estimate of the node at the top is found to be nonpositive.

\eat{
A useful optimization is as follows.
When a node is chosen from the top of the heap, before computing its 
benefit, the cost of recomputing it is compared to the cost of reusing it.
If the cost of recomputing it is less, the node is removed from 
consideration, and its benefit need not be computed.
Note that as other nodes get materialized during the execution of the greedy 
algorithm, the cost of recomputing a node decreases; the incremental
cost computation algorithm keeps the cost of computation up-to-date for
all nodes.
}

\fullversion{
As stated earlier, the assumption above that the benefit function that
we use is monotonic is a convenient heuristic. It is possible to
construct cases where this property does not hold true.
%\reminder{
May want to say more about this not occurring too often?
If we have numbers to back up the claim.
%}

%\reminder{
IMPORTANT FOR FUTURE: 
Once a node such as $_A {\cal G}_{aggfn1, aggfn2}(E)$
is chosen for materialization, the nodes with individual aggregations
$aggfn1$ and $aggfn2$ are not likely to be considered since they will 
now be quite inexpensive.
But the other order of choosing them is possible (since the costs will 
be the same).  So we may also want to do some cleaning up where we reevaluate
the benefits of nodes already chosen for materialization, and unmaterialize
them if required.  As a further optimization we only consider unmaterializing
already marked ancestor nodes, when a node gets marked for materialization.

We should put it somewhere down on our priority list for the future
since it adds some depth to our heuristics.  
% }
}

\eat{
If the set of sharable nodes is large, the greedy algorithm could
be inefficient. Let us define the benefit of a node as the  cost of
evaluating \eat{materializing} a node times the degree of sharing
of the node. 
We can now prune the number of sharable nodes
by considering only those whose benefit is in the top $n$;  $n$ is
a parameter that is set based on how long we are willing to let the
optimizer run.  A DBA can set $n$ appropriately based on experience.
}

%%%%%%%%%%%
% Based on past experience of how long the algorithm 
% takes for a certain $n$ for a queries of a certain size can set 
%%%%%%%%%%%

\eat{
 The set of nodes to be considered for sharing is
first pruned by removing from consideration those nodes that satisfy either of
the following conditions :
\begin{enumerate}
\item{Cost of the node is more than the total cost of the initial plan.
Obviously, this node cannot be a part of the final plan.}
\item{The maximum effective cost of the node if less than a given percent of the
total initial cost of the query. This cut--off percentage can be tuned.
The maximum effective cost of a node is calculated
as the product of it's basic cost and the maximum degree of sharing of the node.
The idea behind this pruning is that nodes of insignificant cost can be 
neglected.}
\end{enumerate}

% \reminder{Eaten up some stuff about pruning some sharable nodes  --
% see whether it is really required -- shall we add the sorted 
% sharable node list idea?}
}

\eat{
\subsection{Heuristic 2 : The More Greedy Approach}

 As the name suggests, this heuristic improves on the Greedy Approach 
described above. As before, the set of 
sharable nodes to be considered is pruned, and a subset is
chosen for materialization.

 However, it uses a different approach for
choosing the next node to be materialized. The nodes are sorted according 
to the benefit of materializing the node individually in the initial DAG.
At each step, the node chosen is the one at the head of this sorted list,
as long as it's benefit in the current DAG is at least as much as that of the
node next in the list. If it's benefit has reduced, then it is pushed into
the list and the new node at the head of the sorted list is considered.
Note that the benefit of other nodes in the list may be based on a 
different set of materialized nodes. This benefit is not monotonic. That is,
when the set of materialized nodes is increased, the benefit of a node
can suddenly reduce, and sometimes even become negative.

 The heuristic is terminated when there is no increase in the benefit, or there
are no more nodes to consider. The number of updates required to the DAG 
is quadratic in the number of sharable nodes considered for materialization in
the worst case, and linear in the best case. In general, it is faster
than Heuristic 1, and returns equally good plans.

\subsection{Heuristic 3 : The Reverse More Greedy Approach}
 
   This heuristic adopts a reverse approach while deciding the set
of sharable nodes that should be materialized. It initially assumes that
all sharable nodes are materialized, and then decides which nodes should
be unmaterialized so that the query processing cost is minimized.

 The initial plan is thus found on the query DAG where all
shared nodes are marked as materialized. The basic pruning employed
in the first two heuristics can also be employed here to prune the set of
sharable nodes that will be examined. In fact, the algorithm now works exactly
as in Heuristic 2, except that the nodes are chosen to be unmaterialized. 

 The heuristic is terminated when there is no increase in the benefit, or there
are no more nodes to consider. The number of updates required to the DAG
is again quadratic in the number of sharable nodes considered for
materialization in the worst case, and linear in the best case. 
}

%%%%%%%%%%%%%%%%%%%%%%%%%%%%%%%%%%%%%%%%%%%%%%%%%%%%%%%%%%%%%%%%%%%%%%%%%%%%%%

%% file: nestedindex.tex
\sections{Extensions}
\label{sec:extnew}

In this section, we briefly outline extensions to 
i) incorporate creation and use of temporary indices, 
ii) optimize nested queries to exploit common sub-expressions
and iii) optimize multiple invocations of
parameterized queries.

Costs may be substantially reduced by creating (temporary) indices on 
database relations or materialized intermediate results.   
To incorporate index selection, we model the presence of an index 
as a physical property, similar to sort order.  
Since our algorithms are actually executed on the physical DAG, they choose
not only what results to materialize but also what physical properties
they should have.  Index selection then falls out as simply a special
case of choosing physical properties, with absolutely no changes to our
algorithms.

\fullversion{
Note that our framework allows us to consider materialization of
indices even if the corresponding relation is not materialized, which 
is useful for algorithms such as index-only joins.
}

Next we consider nested queries.  One approach to handling nested
queries is to use decorrelation techniques (see, e.g. \cite{praveen96:decorr}).
The use of such decorrelation techniques results in the query being
transformed to a set of queries, with temporary relations being created.
Now, the queries generated by decorrelation have several subexpressions
in common, and are therefore excellent candidates for multi-query 
optimization.  One of the queries in our performance evaluation brings
out this point.

Correlated evaluation is used in other cases, either because it may
be more efficient on the query, or because it may not be possible
to get an efficient decorrelated query using standard relational 
operations \cite{rao:ross}. 
In correlated evaluation, the nested query is repeatedly invoked  
with different values for correlation variables.
Consider the following query.
\begin{verbatim}
Query: select * from a, b, c
       where a.x = b.x and b.y = c.y and
             a.cost = (select min(a1.cost) from  a as a1, b as b1
                       where a1.x = b1.x and b1.y = c.y)
\end{verbatim}
One option for optimizing correlated evaluation of this query is to 
materialize $a \Join b$, and share it with the outer level query and 
across nested query invocations.
An index on $a \Join b$, on attribute $b.y$ is required for efficient 
access to it in the nested query, since there is a selection on $b.y$ 
from the correlation variable.
If the best plan for the outer level query uses the join order
$(a \Join b) \Join c$, materializing and sharing $a \Join b$ 
may provide the best plan.

In general, parts of the nested query that do not depend on the 
value of correlation variables can potentially be shared 
across invocations \cite{rao:ross}. 
We can extend our algorithms to consider such reuse across multiple
invocations of a nested query.
The key intuition is that when a nested query
is invoked many times, benefits due to materialization must be multiplied
by the number of times it is invoked; results that depend on correlation
variables, however, must not be considered for materialization.
%For lack of space we omit details, but note that the extensions have 
%been implemented in our system.
The nested query invariant optimization techniques of 
\cite{rao:ross} then fall out as a special case of ours.

Our algorithms can also be extended to optimize multiple invocations 
of parameterized queries.
Parameterized queries are queries that take parameter values, which are
used in selection predicates; stored procedures are a common example.
Parts of the query may be invariant, just as in nested queries, 
and these can be exploited by multi-query optimization.

These extensions have been implemented in our system; details 
may be found in \cite{rssb:tr}.

\fullversion{
{\em For the benefit of the referees, details are provided in
Appendix~\ref{app:nested}.}
}

\fullversion{
First, we create the expanded DAG for the nested query, and add it
to the query DAG, but as a part of the selection condition in the outer 
level query.  The outer level query DAG is expanded as before.
Second, any node in the expanded nested query DAG that has a selection 
using the correlation variable, as well as any of its ancestors,
is marked as not suitable for materialization, since its result varies 
across invocations. 
Finally, when computing the benefit of a set of materialized queries,
we find the benefit to the nested query.  At each place in the
DAG where the selection involves the nested query, the benefit is
multiplied by the number of times the nested query is invoked.
The greedy algorithm is unchanged otherwise.

Note that the technique of sharing nested query invariants 
proposed by Rao and Ross \cite{rao:ross} falls out as a special case of
the above.
Decorrelation of nested queries \cite{??} is another alternative to evaluate
nested queries. 
When applicable, generates multiple queries with common subexpressions;  
multi-query optimization of the decorrelated query is clearly a good idea.

\fullversion{
Extensions that take into account memoization of nested query results, 
or even intermediate results of subqueries (along, of course,
with the correlation variable values) are also possible; we omit them 
for lack of space.
}
}

%%%%%%%%%%%%%%%%%%%%%%%%%%%%%%%%%%%%%%%%%%%%%%%%%%%%%%%%%%%%%%%%%%%%%%%%%%%%%%

%% file: perf.tex
\eat{\small \renewcommand{\baselinestretch}{1.0}
{\bf \noindent  The following paragraph is a note for the referees and not
part of the paper} 

Due to lack of space (given the very strict guidelines on paper length), 
we have been forced to omit details of (i) the extension to the DAG generation 
algorithm to perform unification and subsumption, (ii) incremental 
computation on the physical DAG, (iii) the algorithm for efficient 
sharability computation, and (iv) details of how to handle nested queries.
These are important and non-trivial aspects of the multi-query optimization 
problem, which we have solved and implemented on our system, and
the results in Section~\ref{sec:perf} are based on this implementation.  
The above details are available in the full version of the 
paper~\cite{rssb:tr} available on the web at 
http://www.cse.iitb.ernet.in/\textasciitilde prasan/mqo-full.ps.gz
}

\sections{Performance Study}
\label{sec:perf}

\fullversion{
In this section we present a performance study of the algorithms for
multi-query optimization presented in the paper.  The focus is on
determining (a) the benefits and costs of multi-query optimization and (b)
the effects of the various optimizations which we have proposed. 
}

Our algorithms were implemented by extending and modifying a
Volcano-based query optimizer we had developed earlier.  All coding was
done in C++, with the basic optimizer taking approx.\ 17,000 lines,
common MQO code took 1000 lines, Volcano-SH and Volcano-RU took around
500 lines each, and Greedy took about 1,500 lines.

The optimizer rule set consisted of select push down, join commutativity and 
associativity (to generate bushy join trees),
and select and aggregate subsumption.
Our implementation incorporates the optimizations of \cite{pel97:com} which,
for join transformations, prevent repeated derivations of the same expressions.

Implementation algorithms included sort-based aggregation, 
merge join, nested loops join, indexed join, indexed select and 
relation scan. Our implementation incorporates all the
techniques discussed in this paper, including handling 
physical properties (sort order and presence of indices)
on base and intermediate relations, unification and subsumption
during DAG generation, and the sharability algorithm
for the greedy heuristic. 
\fullversion{
We used a catalog which stored statistics for each relation,
such as number of tuples, size of tuples, and number of distinct values
per column in a relation.  Cost estimates were based on standard I/O cost
estimation techniques such as in \cite{sks97:dbconcepts}. 
}

The block size was taken as 4KB and our cost 
functions assume 6MB is available to each operator
during execution (we also conducted experiments with larger
memory sizes up to 128 MB, with similar results).
Standard techniques were used for estimating
costs, using statistics about relations.  The cost estimates contain an
I/O component and a CPU component, with seek time as 10 msec, transfer
time of 2 msec/block for read and 4 msec/block for write, and CPU cost
of 0.2 msec/block of data processed.  \fullversion{ We assume that
intermediate results are not put onto the disk in between successive
operators -- this eliminates the input reading costs for all the
operators.  Results that are shared are assumed to be materialized.  }
We assume that intermediate results are pipelined to the next input,
using an iterator model as in Volcano;  they are saved to disk only if
the result is to be materialized for sharing.  The materialization cost
is the cost of writing out the result sequentially.

The tests were performed on a single processor 233 Mhz Pentium-II machine
with 64 MB memory, % and 130 MB of swap space, 
running Linux.  
Optimization times are measured as CPU time (user+system).

\subsections{Basic Experiments}

The goal of the basic experiments was to quantify the benefits and
cost of the three heuristics for multi-query optimization, Volcano-SH,
Volcano-RU and Greedy, with plain Volcano-style optimization as the base case.
We used the version of Volcano-RU which considers the forward and reverse 
orderings of queries to find sharing possibilities, and chooses the 
minimum cost plan amongst the two.

%% \reminder{ TODO:  Our numbers here are a bit of a lie, since we 
%%      do not measure the time -- we run the whole thing twice and find
%%      min cost, and min time(!), whereas we should be changing the RU code 
%%      and finding correct time.  The actual time should be very close
%%      since most of the time would be spent in DAG generation which is
%%      done only once. }

%% The same transformation rules and cost model were used 
%% for all the alternatives.

\begin{figure*}[t]
\centerline{
    %\mbox{\pdfimage width 3.0in {graphs/tpcd-cost.pdf} \relax}~~
    %\mbox{\pdfimage width 3.0in {graphs/tpcd-time.pdf} \relax}
     \psfig{file=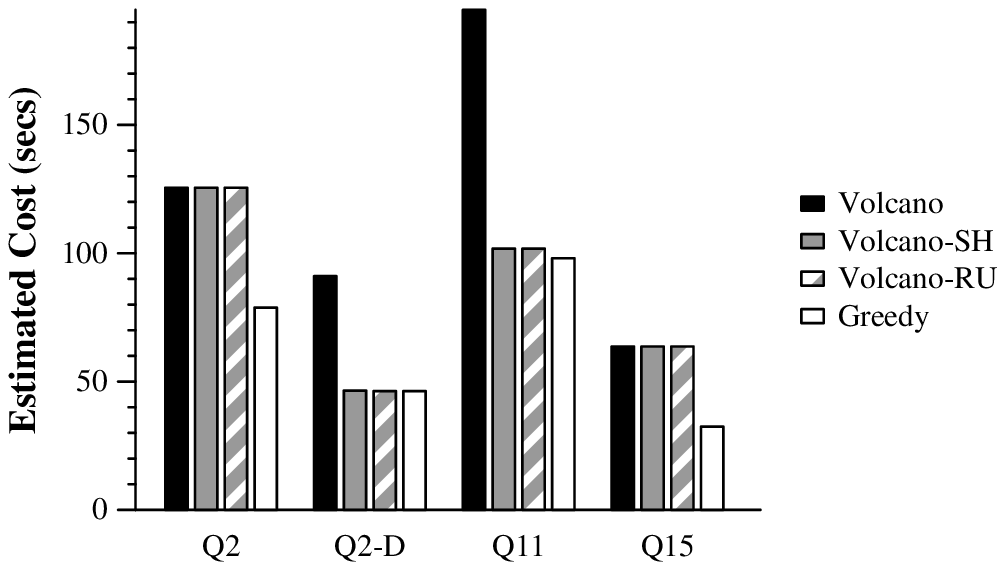,width=3.0in,height=2in} ~~
     \psfig{file=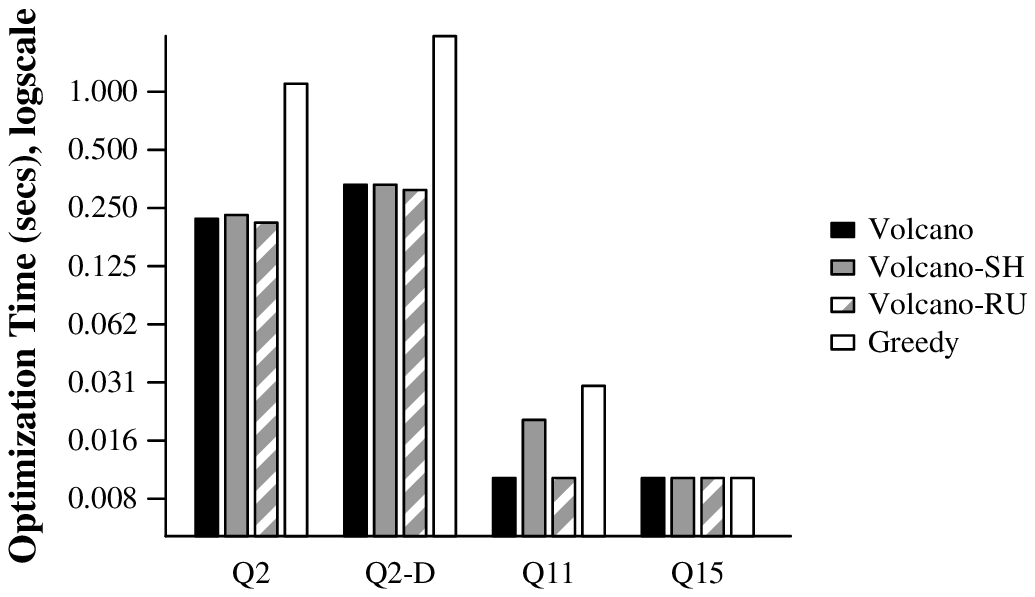,width=3.0in,height=2in}
}
\caption{Optimization of Stand-alone TPCD Queries}
\label{fig:tpcd}
\end{figure*}

\itemhead{Experiment 1 (Stand-Alone TPCD)}
The workload for the first experiment consisted of four queries based on
the TPCD benchmark.
We used the TPCD database at scale of 1 (i.e., 1 GB total size), with 
a clustered index on the primary keys for all the base relations.
The results are discussed below and plotted in Figure~\ref{fig:tpcd}.

TPCD query Q2 has a large nested query, and repeated invocations of the 
nested query in a correlated evaluation could benefit from reusing some 
of the intermediate results.
For this query, though Volcano-SH and Volcano-RU do not lead to any
improvement over the plan of estimated cost 126 secs.\ returned by
Volcano, Greedy results in a plan of with significantly reduced cost
estimate of 79 secs.
Decorrelation is an alternative to correlated evaluation, and 
Q2-D is a (manually) decorrelated version of Q2 (due to decorrelation,
Q2-D is actually a batch of queries).
%%%%%%%%%%%
% For Q2-D, Volcano arrived at a plan with estimated cost of 91 secs, 
% which is significantly more than the estimated time of 79 secs.\ for 
% the plan obtained by Greedy on undecorrelated Q2! 
% This very clearly brings out the case for MQO.
%%%%%%%%%%%
Multi-query optimization also gives substantial gains on the decorrelated 
query Q2-D, resulting in a plan with estimated costs of 46 secs.,
since decorrelation results in common subexpressions.
Clearly the best plan here is multi-query optimization coupled
with decorrelation.  

Observe also that the cost of Q2 (without decorrelation) with Greedy
is much less than with Volcano, and is less than even the cost of Q2-D 
with plain Volcano --- this results indicates that multi-query 
optimization can be very useful in other queries where decorrelation is 
not possible.
To test this, we ran our optimizer on a variant of Q2 where the {\sf in} 
clause is changed to {\sf not in} clause, which prevents decorrelation 
from being introduced without introducing new internal operators such 
as anti-semijoin \cite{rao:ross}.
We also replaced the correlated predicate $PS\_PARTKEY = P\_PARTKEY$ by 
$PS\_PARTKEY \not = P\_PARTKEY$.  For this modified query, Volcano 
gave a plan with estimated cost of 62927 secs., while Greedy was 
able to arrive at a plan with estimated cost 7331, an improvement by 
almost a factor of 9.

We next considered the TPCD queries Q11 and Q15, both of which have
common subexpressions, and hence make a case for multi-query 
optimization.\footnote{As mentioned earlier, we use the term multi-query
optimization to mean optimization that exploits common subexpressions,
whether across queries or within a query.} 
\eat{
While Volcano-SH and Volcano-RU do not lead to appreciable improvements,
%% \reminder{Sud: IMP: Any idea why?  We shoudl explain...}
Greedy is able to detect the common subexpressions and arrives at
plans of approximately half the cost as those returned by Volcano,
as can be seen from Figure~\ref{fig:tpcd}.
% \reminder{IMPIMP: fix with new figures if any...}
}
For Q11, each of our three algorithms lead to a plan of approximately
half the cost as that returned by Volcano. Greedy arrives at similar
improvements for Q15 also, but Volcano-SH and Volcano-RU do not lead to
any appreciable benefit for this query.

% while Volcano results in plans of estimated times 195 secs.\ and 64
% secs.\ for Q11 and Q15 respectively, Greedy arrives at plans of
% estimated times 102 secs.\ and 32 secs.\ respectively.

\eat{
By our cost estimates, substantial gains are obtained due to MQO on the
queries considered. Most significantly, we observed that the estimated
cost of the MQO plan for uncorrelated Q2 obtained by Greedy is much
less than the estimated cost of the plan obtained by Volcano for the
decorrelated version Q2-D. Moreover, Volcano-RU and Greedy detect the
common subexpression in the decorrelated query Q2-D and bring out even
further reductions in the cost as a result. The plan returned by
Volcano for query Q2 has a cost estimate of 126 secs. In contrast,
Greedy returns a plan with cost estimate of 79 secs.  Applying
decorrelation on Q2 results in Q2-D for which Volcano return a plan
with estimated cost of 91 secs.  Applying Greedy on Q2-D results in a
MQO plan with estimated cost of 46 secs.
}

Overall, Volcano-SH and Volcano-RU take the same time and space as Volcano.
Greedy takes more time than the others for all the queries, but 
the maximum time taken by greedy over the four queries was just under 2 
seconds, versus 0.33 seconds taken by Volcano for the same query.
The extra overhead of greedy is negligible compared to its benefits.
% For Q11 and Q15, the times taken by Volcano were practically
% at the minimum of our measurement resolution (leading to some measurement
% noise, where Volcano-RU appears to run faster than Volcano).
% In terms of relative time taken, Greedy needed a maximum of about 6
% times as much time as Volcano, but took a maximum of just under 2 seconds, 
% which is very small compared to its benefits.
\eat{
In fact, in the figure, some bars are not visible since the corresponding 
times were very small (less than 0.01 secs.).
}
The total space required by Greedy ranged from 1.5 to 2.5 times that of
the other algorithms, and again the absolute values were quite small
(up to just over 130KB).

\noindent{\bf Results on Microsoft SQL-Server 6.5:}

\begin{figure*}[t]
\centerline{
    %\mbox{\pdfimage width 3.0in {graphs/mssql-cost.pdf} \relax}
     \psfig{file=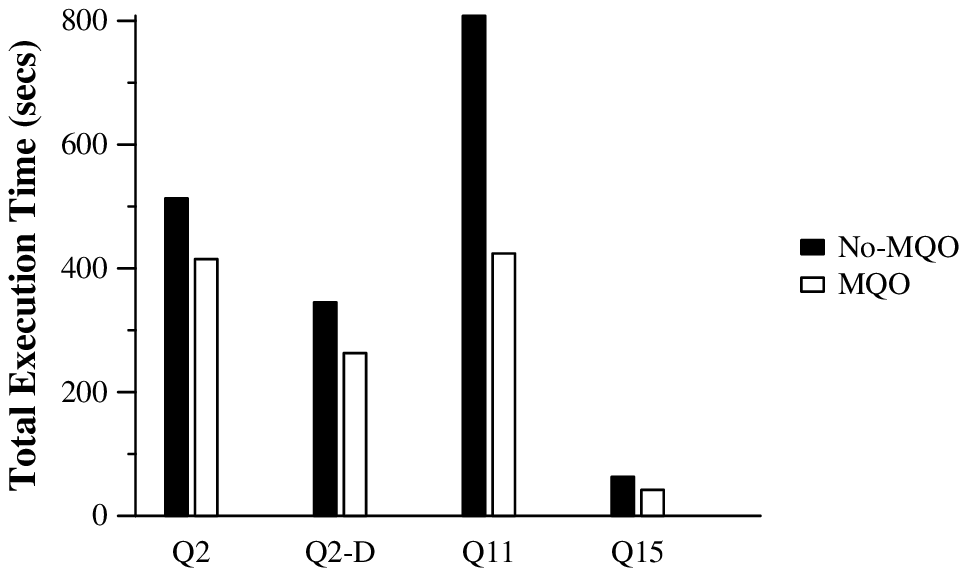,width=3.0in,height=2in}
}
\caption{Execution of Stand-alone TPCD Queries on MS SQL Server}
\label{fig:mssql-tpcd}
\end{figure*}

To study the benefits of multi-query optimization on a real database, 
we tested its effect on the queries mentioned above, 
executed on Microsoft SQL Server 6.5, running on Windows NT, 
on a 333 Mhz Pentium-II machine with 64MB memory.  
We used the TPCD database at scale 1 for the tests.
To do so, we encoded the plans generated by Greedy into SQL.
We modeled sharing decisions by creating
temporary relations, populating, using and deleting them.  If so indicated by
Greedy, we created indexes on these temporary relations. 
We could not encode the exact evaluation plan in SQL since SQL-Server 
does its own optimization. 
We measured the total elapsed time for executing all these steps.  

The results are shown in Figure~\ref{fig:mssql-tpcd}.
For query Q2, the time taken reduced from 513 secs.\ to 415 secs.  
Here, SQL-Server performed decorrelation on the original Q2 as well as on
the result of multi-query optimization.  Thus, the numbers do not match
our cost estimates, but clearly multi-query optimization was useful here.
The reduction for the decorrelated version Q2-D was from 345 secs.\ to 
262 secs; thus the best plan for Q2 overall, even on SQL-Server,
was using multi-query optimization as per Greedy on a decorrelated query.
The query Q11 speeded up by just under 50\%, from 808 secs.\ to 
424 secs.\  and Q15 from 63 secs.\ to 42 secs.\ using plans with sharing
generated by Greedy.
\fullversion{
Query QR2, which is not decorrelatable \cite{rao:ross}, also showed 
an improvement from 36 secs.\ to 28 secs.  
}

The results indicate that multi-query optimization gives significant
time improvements on a real system.
It is important to note that the measured benefits are 
underestimates of potential benefits,
for the following reasons.
(a) Due to encoding of sharing in SQL, temporary relations had 
to be stored and re-read even for the first use.  
If sharing were incorporated within the evaluation engine, the 
first (non-index) use can be pipelined, reducing the cost further.
(b) The operator set for SQL-Server 6.5
\reminder{Prasan: The version I am looking at has sort-merge joins}
seems to be rather restricted, and
does not seem to support sort-merge join; for all queries we submitted, 
it only used (index)nested-loops. Our optimizer at times indicated that 
it was worthwhile to materialize the relation in a sorted order so 
that it could be cheaply used by a merge-join or aggregation over it,
which we could not encode in SQL/SQL-Server.

In other words, if multi-query optimization were properly integrated 
into the system, the benefits are likely to be significantly larger, 
and more consistent with benefits according to our cost
estimates.
\eat{
\footnote{Work is ongoing at Microsoft Research to prototype
multiquery optimization in Microsoft SQL-Server.}
\reminder{Sud:  Prasan, pl check with Paul if its OK to have the above
footnote.}
\reminder{Prasan: Paul says it is too early to make this public}
}

\eat{
\noindent{\bf Overheads of Multi-Query Optimization}

While the benefits of using MQO show up on query workloads with common
subexpressions, a relevant issue is the performance on workloads
with rare or nonexistent overlaps.
If it is known apriori that the workload is not going to benefit
from MQO, then we can set a flag in our optimizer
that bypasses the MQO related algorithms described in this paper,
reducing to plain Volcano.

However, if the workload characteristics are not known apriori then it
is important that the MQO does not pose excessive overheads on queries
which do not benefit from it. We therefore studied the overheads of our
algorithms on queries Q3, Q5, Q7, Q9 and Q10 from TPCD; none of these
queries has any common subexpressions.

For each query, the plan returned by each of the algorithms was identical,
as expected.
The overheads of MQO are negligible for Volcano-SH and Volcano-RU, whuch
is expected since these algorithms are designed as lightweight extensions
to Volcano.  In the case of Greedy, the sharability
detection algorithm (described in Section~\ref{app:sha}) finds no node
sharable, causing the Greedy algorithm to terminate immediately. Thus,
the only overhead in Greedy is the execution of the sharability detection
algorithm, which accounts for at most 60\% overhead over plain Volcano.

\begin{figure*}[t]
\centerline{
    %\mbox{\pdfimage width 3.0in {graphs/nomqo-time.pdf} \relax}
     \psfig{file=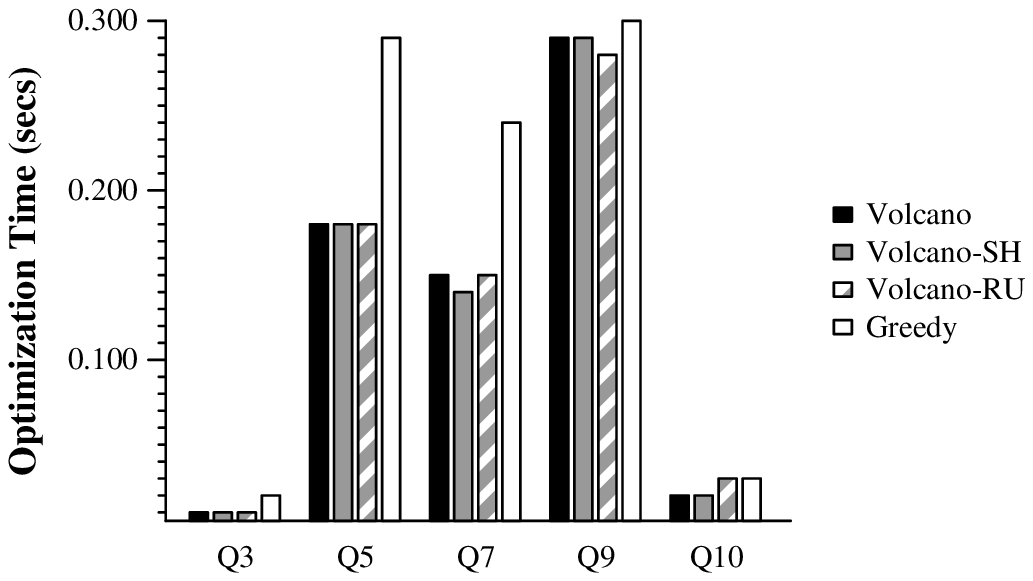,width=3.0in,height=2in}
}
\caption{Optimization of TPCD Queries without Common Subexpressions}
\label{fig:nomqo}
\end{figure*}
}
\itemhead{Experiment 2 (Batched TPCD Queries)}
In the second experiment, the workload models a system where several
TPCD queries are executed as a batch.  The workload consists of
subsequences of the queries Q3, Q5, Q7, Q9 and Q10 from TPCD; none
of these queries has any common subexpressions within itself.  Each
query was repeated twice with different selection constants.  
Composite query BQi consists of the first i of the above queries, and
we used composite queries BQ1 to BQ5 in our experiments.  
Like in Experiment 1, we used the TPCD database at scale of 1 and 
assumed that there are clustered indices on the primary keys of the 
database relations.

Note that although a query is repeated with two different values for
a selection constant, we found that the selection operation 
generally lands up at the bottom of the best Volcano plan tree, 
and the two best plan trees may not have common subexpressions.
\reminder{recheck this}

\begin{figure*}[t]
\centerline{
    %\mbox{\pdfimage width 3.0in {graphs/tpcd-comp-cost.pdf} \relax} ~~
    %\mbox{\pdfimage width 3.0in {graphs/tpcd-comp-time.pdf} \relax}
     \psfig{file=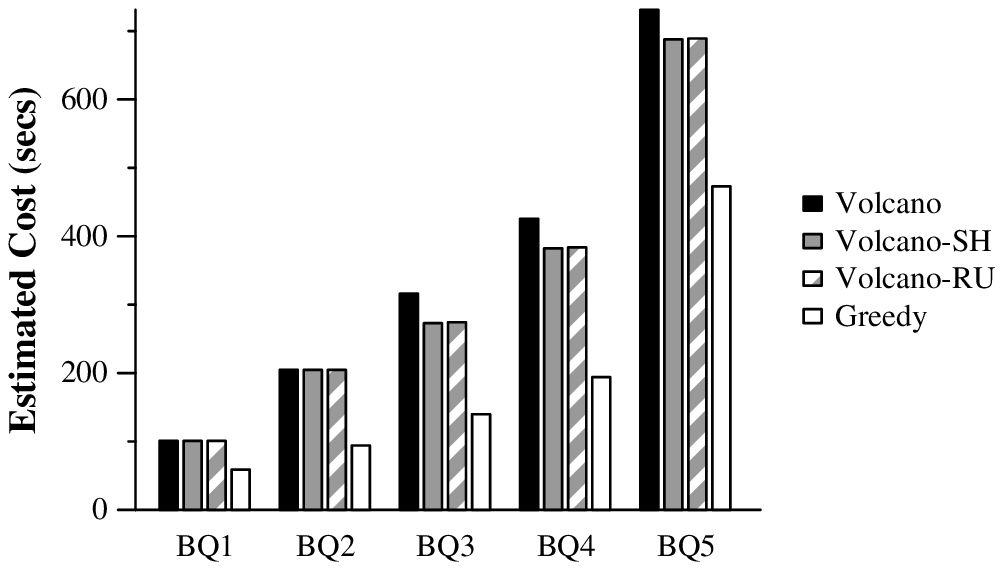,width=3.0in,height=2in} ~~
     \psfig{file=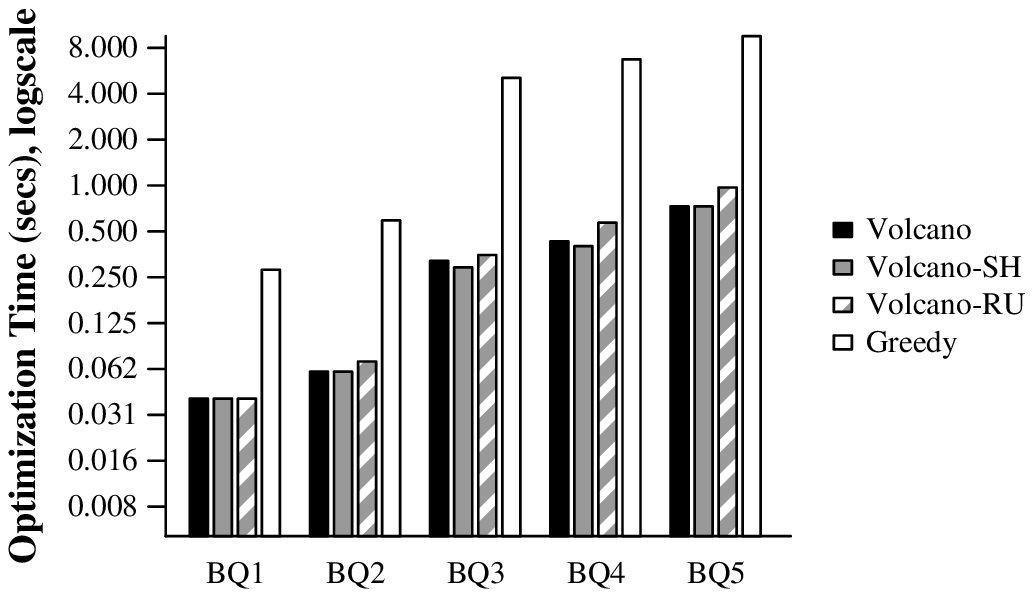,width=3.0in,height=2in}
}
\caption{Optimization of Batched TPCD Queries}
\label{fig:tpcd-comp}
\end{figure*}

The results on the above workload are shown in
Figure~\ref{fig:tpcd-comp}.  Across the workload, Volcano-SH and
Volcano-RU achieve up to only about 14\% improvement over Volcano with
respect to the cost of the returned plan, while incurring negligible
overheads.  There was no difference between Volcano-SH and Volcano-RU
on these queries, implying the choice of plans for earlier queries
did not change the local best plans for later queries.
Greedy performs better, achieving up to 56\% improvement
over Volcano, and is uniformly better than the other two algorithms.

As expected, Volcano-SH and Volcano-RU have essentially the same execution 
time and space requirements as Volcano.  
Greedy takes about 10 seconds on the largest query in the set,
BQ5, while Volcano takes about 0.7 second on the same.
However, the estimated cost savings on BQ5 is 260 seconds, which is
clearly much more than the extra optimization time cost of 10 secs.
Thus the extra time spent on Greedy is well spent.  Similarly, the
space requirements for Greedy were more by about a factor of three to
four over Volcano, but the absolute difference for BQ5 was only 60KB.  The
benefits of Greedy, therefore, clearly outweigh the cost.

\reminder{Sud: should we explain why we have no SQL server numbers??}

\subsections{Scaleup Analysis}

To see how well our algorithms scale up with increasing numbers of
queries, we defined a new set of 22 relations $PSP_1$ to $PSP_{22}$
with an identical schema $(P$, $SP$, $NUM)$ denoting part id, subpart
id and number.
Over these relations, we defined a sequence of 18
component queries $SQ_1$ to $SQ_{18}$: component query $SQ_i$ was a
pair of chain queries on five consecutive relations $PSP_i$ to
$PSP_{i+4}$, with the join condition being $PSP_j.SP = PSP_{j+1}.P$,
for $j=i..i+3$.  
One of the queries in the pair $SQ_i$ had a selection $PSP_i.NUM \geq a_i$ 
while the other had a selection $PSP_i.NUM \geq b_i$ where $a_i$ and $b_i$ 
are arbitrary values with $a_i \not = b_i$.

To measure scaleup, we use the composite queries $CQ_1$ to
$CQ_{5}$, where $CQ_i$ is consists of queries $SQ_1$ to $SQ_{4i-2}$.
Thus, $CQ_i$ uses $4i+2$ relations $PSP_1$ to $PSP_{4i+2}$, and has
$32i-16$ join predicates and $8i-4$ selection predicates.
Query CQ5, in particular, is on 22 relations and has 144
join predicates and 36 select predicates.
The size of the 22 base relations $PSP_1, \ldots, PSP_{22}$ varied from
20000 to 40000 tuples (assigned randomly) with 25 tuples per block.
No index was assumed on the base relations. 

\begin{figure*}
\centerline{
    %\mbox{\pdfimage width 3.0in {graphs/stag-cost.pdf} \relax} ~~
    %\mbox{\pdfimage width 3.0in {graphs/stag-time.pdf} \relax}
     \psfig{file=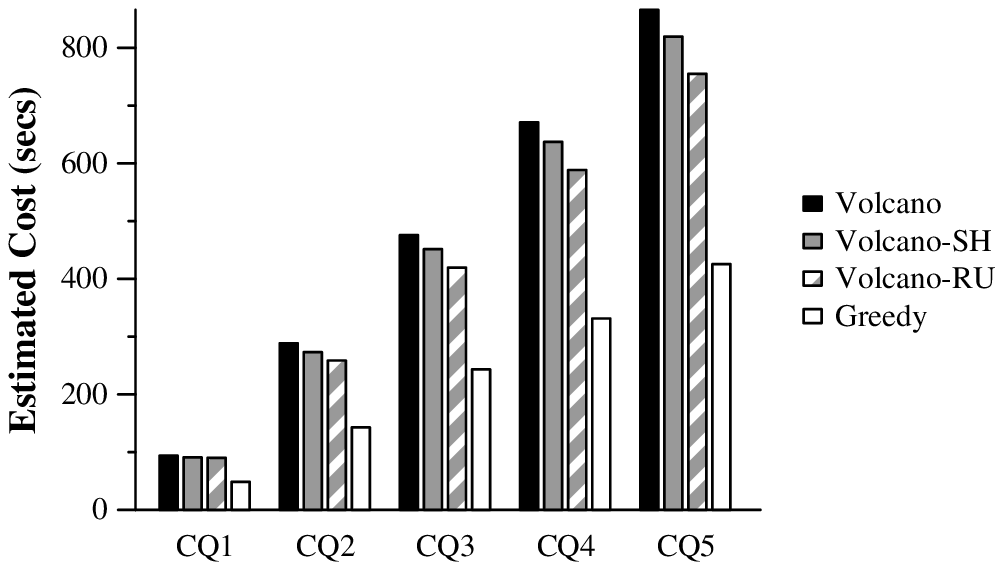,width=3.0in,height=2in} ~~
     \psfig{file=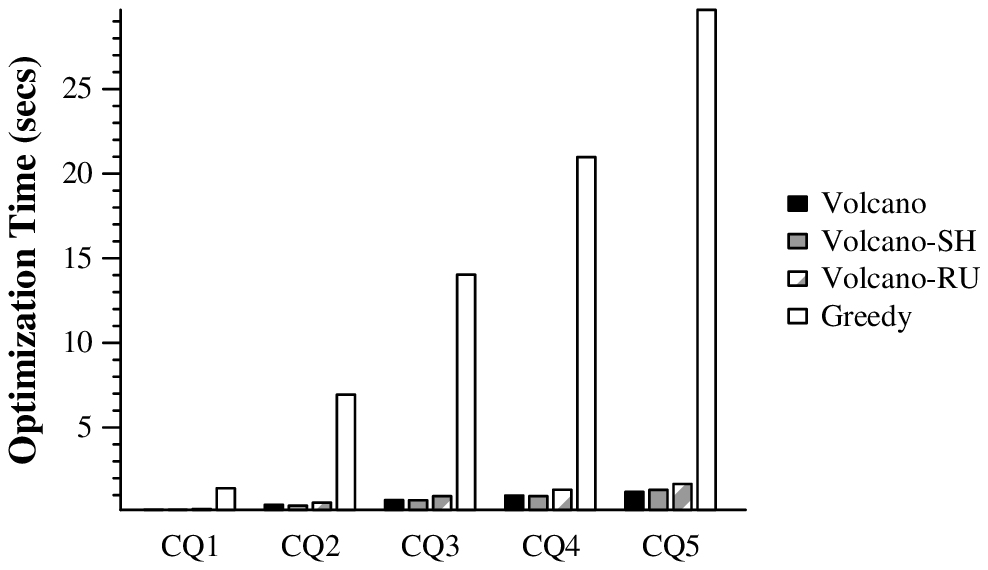,width=3.0in,height=2in}
}
\caption{Optimization of Scaleup Queries}
\label{fig:stag}
\end{figure*}

\eat{
%%
%% all this was observed in expt 2 -- no point repeating the same thing here
%%
The results show that for each of these queries, only Greedy finds out
a plan that effectively shares the common subexpressions leading to
savings of up to 35\% over the Volcano plan across the range of
queries.  The savings with Volcano-SH and Volcano-RU are substantially
smaller at about 10\%.  For query CQ5, which is on 22 relations and has 144
join predicates and 36 select predicates, Greedy takes
just an additional 31 seconds over Volcano.
}

The cost of the plan and optimization time for the above workload is
shown in Figure~\ref{fig:stag}.
The relative benefits of the algorithms remains similar to that in
the earlier workloads, except that Volcano-RU now gives somewhat better
plans than Volcano-SH.  Greedy continues to be the best, although it
is relatively more expensive.
The optimization time for Volcano, Volcano-SH and Volcano-RU increases
linearly.  The increase in optimization time for Greedy is also practically 
linear, although it has a very small super-linear component.  
But even for the largest query, CQ5 (with 22 relations, 144 join 
predicates and 36 select predicates) the time taken was only 30 seconds.
The size of the DAG increases linearly for this sequence of queries.
From the above, we can conclude that Greedy is scalable to quite large
query batch sizes.

\eat{
%%
%% significant, but out of context here
%%
The absolute time and space
utilization indicate that Greedy is quite cheap, even for reasonably
large query sets.  Greedy has a slightly higher space utilization than
the others, but even for a query as big as CQ5, Greedy requires only
1.7MB space while Volcano requires 0.6MB space.
}

\begin{figure*}
\centerline{
    %\mbox{\pdfimage width 3.0in {graphs/stag-propcount.pdf} \relax} ~~
    %\mbox{\pdfimage width 3.0in {graphs/stag-numprops.pdf} \relax}
     \psfig{file=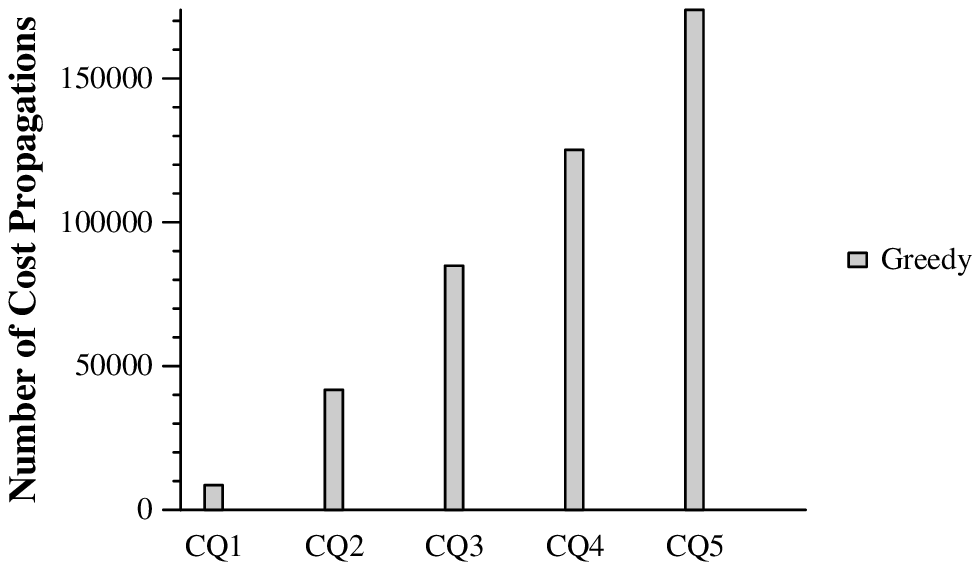,width=3.0in} ~~
     \psfig{file=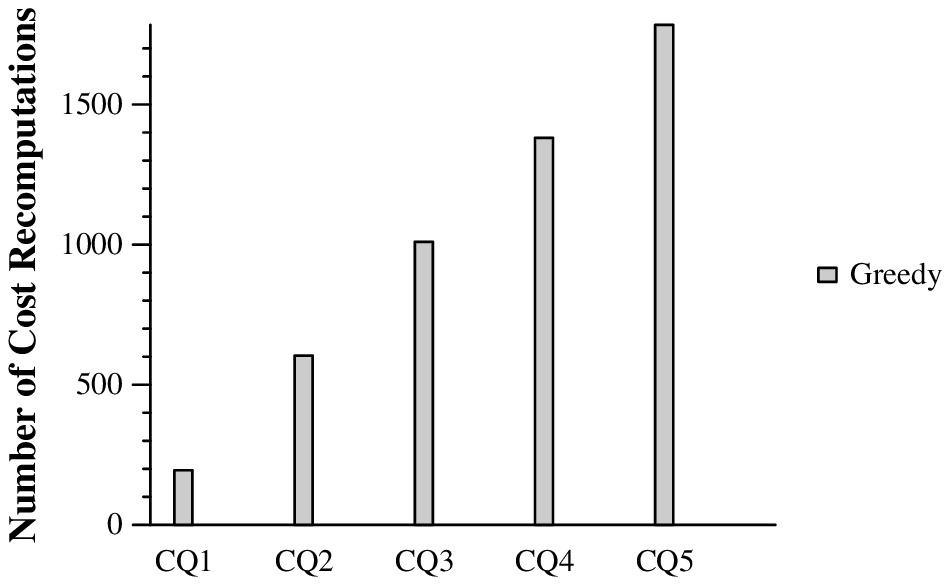,width=3.0in}
}
\caption{Complexity of the Greedy Heuristic}
\label{fig:Greedy-complexity}
\end{figure*}

To better understand the complexity of the Greedy heuristic on the
scaleup workload, in addition to the optimization time we measured the
total number of times cost propagation occurs across equivalence nodes,
and the total number of times cost recomputation is initiated.  The
result is plotted in Figure~\ref{fig:Greedy-complexity}.  Note that in
addition to the  size of the DAG, the number of sharable nodes
also increases linearly across queries CQ1 to CQ5.

Greedy was considered expensive by \cite{wisc:matview} 
because of its worst case complexity: it can be as much as $O(k^2e)$, 
where $k$ is the number of nodes in the DAG which are sharable,
and $e$ is the number of edges in the DAG.
However, for multi-query optimization, the DAG tends to be wide rather
than tall -- as we add queries, the DAG gets wider, but its height does
not increase, since the height is defined by individual queries.

The result shows that for the given workload,
the number of times cost propagation occurs across 
equivalence nodes, and the number of times cost recomputation is initiated
both increase almost linearly with number of queries.
The observed complexity is thus much less than the worst case complexity.

The number of times costs are propagated across equivalence nodes is 
almost constant per cost recomputation.
This is because the number of nodes of the DAG affected by a 
single materialization does not vary much with number of queries,
which is exploited by incremental cost recomputation.
The height of the DAG remains constant (since the number of relations
per query is fixed, which is a reasonable assumption).  

\subsections{Effect of Optimizations}

In this series of experiments, we focus on the effect of individual
optimizations on the optimization of the scaleup queries.  We first
consider the effect of the monotonicity heuristic addition to Greedy.
Without the monotonicity heuristic, before a node is materialized the
benefits would be recomputed for all the sharable nodes not yet
materialized.  With the monotonicity heuristic addition, we found that
on an average only about 45 benefits were recomputed each time, across
the range of CQ1 to CQ5.  In contrast, without the monotonicity
heuristic, even at CQ2 there were about 1558 benefit recomputations
each time, leading to an optimization time of 77 seconds for the query,
as against 7 seconds with monotonicity. Scaleup is also much worse
without monotonicity.  Best of all, the plans produced with and without
the monotonicity heuristic assumption had virtually the same cost on
the queries we ran.
Thus, the monotonicity heuristic provides very large
time benefits, without affecting the quality of the plans generated.
% , indicating the monotonicity assumption does hold on our benchmark.  

To find the benefit of the sharability computation, we measured the
cost of Greedy with the sharability computation turned off; every node
is assumed to be potentially sharable.  Across the range of scaleup
queries, we found that the optimization time increased significantly.
For CQ2, the optimization time increased from 30 secs.\ to 46 secs.
\eat{
For BQ5, the optimization time increased
14 secs.\ to 19 secs.
}
Thus, sharability computation is also a very useful optimization.

In summary, our optimizations of the implementation of the greedy heuristic
result in an order of magnitude improvement in its performance, 
and are critical for it to be of practical use.

\subsections{Discussion}
To check the effect of memory size on our results, we
ran all the above experiments increasing the memory available to 
the operators from 6MB to 32MB and further to 128MB.  
We found that the cost estimates for the plans decreased slightly, but 
the relative gains (i.e., cost ratio with respect to Volcano)
essentially remained the same throughout for the different heuristics.

We stress that while the cost of optimization is independent of the
database size, the execution cost of a query, and hence the benefit due to
optimization, depends upon the size of the underlying data.
Correspondingly, the benefit to cost ratio for our algorithms increase 
markedly with the size of the data.
To illustrate this fact, we ran the batched TPCD query BQ5 (considered in
Experiment 2) on TPCD database with scale of 100 (total size 100GB).
Volcano returned a plan with estimated cost of 106897 seconds while Greedy
obtains a plan with cost estimate 73143 seconds, an improvement of 33754
seconds. The extra time spent during optimization is 10 seconds, as
before, which is negligible relative to the gain.

%% For the query described above, Volcano-SH and Volcano-RU could
%% obtain only a marginally better plan with estimated cost 62046.
% Smarter optimization makes sense only if the cost incurred during
% optimization is offset by the savings in the execution time for that
% query.  

While the benefits of using MQO show up on query workloads with common
subexpressions, a relevant issue is the performance on workloads with rare
or nonexistent overlaps.  If it is known apriori that the workload is
not going to benefit from MQO, then we can set a flag in our optimizer
that bypasses the MQO related algorithms described in this paper,
reducing to plain Volcano.  
%%%%%%%%%%%%
% However, if the workload characteristics
% are not known apriori then it is important that the MQO does not pose
% excessive overheads on queries which do not benefit from it.
%%%%%%%%%%%%

To study the overheads of our algorithms in a case with no sharing,
we took TPCD queries Q3, Q5, Q7, Q9 and Q10, renamed the relations to
remove all overlaps between queries, and created a batch consisting of 
the queries with relations renamed.
The overheads of Volcano-SH and Volcano-RU are neglibible, as
discussed earlier.  
Basic Volcano optimization took 650 msec, while the Greedy algorithm
took 820 msec.  Thus the overhead was around 25\%, but note that the
absolute numbers are very small.  
With no overlap, the sharability detection algorithm finds no node sharable, 
causing the Greedy algorithm to terminate immediately (returning the 
same plan as Volcano). 
Thus, the overhead in Greedy is due to (a) expansion of the entire DAG, and 
(b) the execution of the sharability detection algorithm.
Of this overhead, cause (a) is predominant, and the sharability computation 
was quite cheap on queries with no sharing.
\reminder{perhaps hide some details above to avoid confusing people?}

In our experiments, Volcano-RU was better than Volcano-SH only in a few
cases, but since their run times are similar, Volcano-RU is preferable. 
There exist cases where Volcano-RU finds out plans as good as Greedy in
a much less time and using much less space; but on the other hand, in
the above experiments we saw many cases where additional investment of time
and space in Greedy pays off and we get substantial improvements in the
plan. 

To summarize, for very low cost queries, which take only a few seconds,
one may want to use Volcano-RU, which does a ``quick-and-dirty'' job;
especially so if the query is also syntactically complex.  For more
expensive queries, as well as ``canned'' queries that are optimized
rarely but executed frequently over large databases, it clearly makes
sense to use Greedy.

%% file: relwork.tex
\sections{Related Work}
\label{sec:related}

\eat{~\cite{rou82:view,rosenthal82} suggest using
an AND-OR DAG  for representing alternate query execution plans.
The AND-OR-DAG is very compact 
since logically equivalent expressions in two different query plans 
are stored only once. 
%The AND-OR-DAG representation
%was later adopted by the Volcano Optimizer generator~\cite{gra:vol}.
}

\fullversion{
%%%%%%%%%%%%%%%%%%%%%%%%%%%%%%%
To systematically contrast our work with previous work, we 
split the multi-query optimization problem into three parts: 
(i) Find what expressions can be shared (either since
they are equivalent or via subsumption), and thus may be worth materializing;
(ii) Given a set of materialized expressions, find the best plan
for a query, with the possibility of reusing the materialized results,
in a cost-based manner; and 
(iii) Find which expressions to materialize in order to minimize
overall costs; the solutions to parts (i) and (ii) are used by
the algorithms for this part.
%%%%%%%%%%%%%%%%%%%%%%%%%%%%%%%
}
% parts: i) choosing a set of expressions to check for 
% potential sharing ii) figuring out whether they can actually be shared
% (matching) iii) figuring out the best plan that uses 
% some of them (search strategy). 

The multi-query optimization problem
has been addressed in~\cite{fink82,tim:mul, 
kyu:imp,sg:tkde90,cls93:multi,joo:usi,rou:edbt94,wisc:mqo,shivku98:transview}.
The work in~\cite{tim:mul, kyu:imp, sg:tkde90, cls93:multi,joo:usi}
describe exhaustive algorithms;  they use an abstract representation of 
a query $Q_i$ as a set of alternative plans $P_{i,j}$, each having 
a set of tasks $t_{i,j,k}$, where the tasks may be shared between 
plans for different queries.
They do not exploit the hierarchical nature of query optimization problems,
where tasks have subtasks.
Finally, these solutions are not integrated with an optimizer. 
%These solutions, like \cite{fink82}, separate the jobs of finding
%what tasks are shared and how to find the best plan.
%As pointed out by \cite{rou:edbt94,surajit:matviewopt}, separating out
%the two stages leads to an extremely large search space; it is better 
%to integrate the search spaces to recognize commonality among tasks, 
%and thereby avoid duplicate effort (as we do, using the DAG representation).

The work in~\cite{shivku98:transview} considers sharing only amongst the
best plans of each query -- this is similar to Volcano-SH, and
as we have seen, this often does not yield the best sharing.  

%The work of \cite{sellis88:caching} also addresses cache management,
%but as mentioned earlier does not address either cost-based optimization
%or the problem of choosing what views to materialize.
% The class of queries we consider contains both of the
% above classes. % 

The problem of materialized view/index selection~\cite{rou82:view,rss96:matview,
ykl97:matview,sn97:indexsel,lqa97:phys,gupta97:viewsel}
is related to the multi-query optimization problem.
The issue of materialized view/index selection for the special case of 
aggregates/data-cubes is considered in \cite{venky:sigmod96,venky:index}
and implemented in Redbrick Vista \cite{vista:icde98}.
The view selection problem can be viewed as finding the
best set of sub-expressions to materialize,
given a workload consisting of both queries and updates.
The multi-query optimization problem differs from the above since it
assumes absence of updates, but it must keep in mind the cost of
computing the shared expressions, whereas the view selection problem
concentrates on the cost of keeping shared expressions up-to-date. 
It is also interesting to note that multi-query optimization is needed
for finding the best way of propagating updates on base relations to 
materialized views~\cite{rss96:matview}.

Several of the algorithms presented for the view selection problem 
(\cite{venky:sigmod96,venky:index,gupta97:viewsel}) are 
similar in spirit to our greedy algorithm, but none of them described how
to efficiently implement the greedy heuristic.
Our major contribution here lies in making the greedy heuristic practical
through our optimizations of its implementation.
We show how to integrate the heuristic with the optimizer, allowing
incremental recomputation of benefits, which was not considered in any
of the earlier papers, and our sharability and monotonicity optimizations
also result in great savings.
The lack of an efficient implementation could be one reason for the 
authors in~\cite{wisc:matview} to claim that the greedy algorithm can 
be quite inefficient for selecting views to materialize for cube queries.
Another reason is that, for multi-query optimization of normal SQL queries
(modeled by our TPC-D based benchmarks) the DAG is ``short and fat'',
whereas DAGs for complicated cube queries tend to be taller.
Our performance study (Section~\ref{sec:perf}) indicates the greedy
heuristic is quite efficient, thanks to our optimizations.

Another related area is that of caching of query results.
Whereas multiquery optimization can optimize a batch of queries given
together, caching takes a sequence of queries over time, deciding what 
to materialize and keep in the cache as each query is processed.
Related work in caching includes \cite{rou:edbt94,wisc:mqo,rou99:dynamat}.
The work in~\cite{wisc:mqo,rou99:dynamat} considers only
queries that can be expressed as a single multi-dimensional expression.
The work in~\cite{rou:edbt94} addresses the issue of 
management of a cache of previous results 
% (which is more general than multi-query optimization in some ways), 
but considers only select-project-join (SPJ) queries.
We consider a more general class of queries.

Our multi-query optimization algorithms implement query optimization
in the presence of materialized/cached views, as a subroutine. By virtue of
working on a general DAG structure, our techniques are extensible, unlike 
the solutions of \cite{surajit:matviewopt} and \cite{rou:edbt94}.
The problem of detecting whether an expression can be used to 
compute another has also been studied in~\cite{larson85,larson87,
sellis88:caching}; however, they do not address the problem of choosing 
what to materialize, or the problem of finding the best query plans 
in a cost-based fashion.

Recently,~\cite{rao:ross} considers the 
problem of detecting invariant parts of a nested subquery,
and teaching the optimizer to choose a plan that
keeps the invariant part as large as possible. 
Performing multi-query optimization on nested queries automatically 
solves the problem they address. 

\eat{   
Our techniques for handling nested queries and
indexes on intermediate relations are useful in the context
of a traditional optimizer too.  Typically, nested queries
have been optimized block at a time, thus not sharing 
optimization effort between blocks. Our technique allows 
the optimization effort to be shared (not repeated).
To the best of our knowledge, no prior work has considered
the problem of creating indexes on intermediate relations
within an optimizer.   
}

Our algorithms have been described in the context of a
Volcano-like optimizer; at least two commercial database systems, 
from Microsoft and Tandem, use Volcano based optimizers.
%%%%%%%%
% As a result of using a Volcano-like framework, our multi-query optimization 
% algorithms can handle new operators easily, and are thus very extensible.
% Our implementation also incorporates the techniques of \cite{pel97:com} 
% to eliminate duplicate derivations during DAG generation, and is
% thus very efficient.
%%%%%%%
However, our algorithms can also be modified to be added on top of
existing System-R style bottom-up optimizers; the main change would be 
in the way the DAG is represented and constructed.

\fullversion{
Thus, they offer solutions to part (ii) above.
However, the solutions do not have to choose what expressions are
to be made available for sharing, since this is fixed a priori
(these are the materialized or the cached views).
Thus, they do not address (iii) above.
In contrast, in multi-query optimization, any expression can be
potentially be chosen to be materialized and shared.
We need to choose the best set for materialization and sharing,
in addition to solving problem (ii) for given sets of materialized
expressions; thus the multi-query optimization problem is more general.
}
\fullversion{
%%%%%%%%%%%%%%%%%%%%%%%
Most of the previous work
on query optimization has been in the context of the System R 
optimizer~\cite{gaclp:acc} primarily because most commercial 
systems used the System R optimizer. 
However, this is changing, and we believe at least two commercial 
One concern about the Volcano optimizer was the number of times 
it would generate the same logical expression (if an expression can
generate $k$ others, it may be generated back $k$ times).
Such duplicate generations do not change the size of the DAG but lead to 
wasted effort.  However, recently it has been shown by ~\cite{pel97:com} that
the transformations can be carefully defined to eliminate
redundant generations of expressions.

Previous work typically either uses the System R join
order optimization algorithm as a subroutine or adds a
preprocessing step to the System R join order optimization algorithm. 
As a result,  several extensions (e.g., common-subexpression elimination
as in \cite{shivku98:transview}, and 
%% optimization in the presence of materialized views~\cite{surajit:matviewopt},
detecting invariant parts of the inner query~\cite{rao:ross}) execute
join-order optimization and the extensions they propose in separate phases. 
Although practical limitations make modifying the optimizer difficult
in some commercial situations, necessitating such solutions,
the price paid is poorer plans and increased optimization time.
%%%%%%%%%%%%%%%%%%%%%%%
}

%% file: conclusions.tex
\sections{Future Work}
\label{sec:future}

The results in this paper form the basis for a significant amount 
of future work.
Our algorithms can be extended to deal with space constraints
on materialized results.
For instance, the greedy algorithm can select equivalence nodes in
order of benefit-per-unit-space, as in \cite{venky:sigmod96,gupta97:viewsel},
until temporary storage space is exhausted.
A more challenging problem is how to schedule computations so that 
temporary storage space can be reused during computation.

Another important area of future work lies in dealing with large sets
of queries (large workloads); the size of the workload can be reduced by
abstracting queries, for instance by replacing queries that differ in
just selection constants by a parameterized query, invoked multiple times.\eat{
Our algorithms can also be extended to deal with physical operations that 
can implement multiple logical operations (such as the cube computation 
algorithms of \cite{prasad:sunita}).
}
Scheduling of multiple pipelines in parallel can help remove the cost 
of materialization in many cases; an example is the shared relation scan
implemented in the Redbrick data warehouse. 
Extending the optimizer to consider such scheduling, including the issue
of partitioning memory amongst the pipelines, is an area we are currently
working on.
%%%%%%%%%%
% The materialization cost can be eliminated or reduced in
% some cases by piggybacking the materialization with the actions
% of an operator that uses the expression.  For instance, if an
% expression is the input to a sort, it can be materialized by 
% simply saving runs generated during sorting, at no extra cost.
%%%%%%%%%%

Another important, and related area, is that of query result caching.
We have recently applied the greedy algorithm presented in this paper
to tackle the problem of cache replacement in query result caching,
reaping substantial benefits.
Details may be found in \cite{rsvrss99:caching} (also submitted
to SIGMOD'2000).
 
We are currently also working on applying multi-query optimization to
incremental update expressions for materialized views.  Initial 
results are very promising.  We also plan to apply these results to
the problem of materialized view/index selection, where update costs need
to be taken into account.

\sections{Conclusions}
\label{sec:conclusions}

We have described three novel heuristic search algorithms, Volcano-SH, 
Volcano-RU and Greedy, for multi-query optimization.
We presented a  a number of techniques to greatly speed up the 
greedy algorithm.
Our algorithms are based on the AND-OR DAG representation of queries, and
are thereby can be easily extended to handle new operators.
Our algorithms also handle index selection and nested queries, in a 
very natural manner. 
We also developed extensions to the DAG generation algorithm to 
detect all common sub expressions and include subsumption derivations. 

Our implementation demonstrated that the algorithms can be added to an
existing optimizer with a reasonably small amount of effort.
Our performance study, using queries based on the TPC-D benchmark, 
demonstrates that multi-query optimization is practical and gives 
significant benefits at a reasonable cost.  The benefits of 
multi-query optimization were also demonstrated on a real database system. 
The greedy strategy uniformly gave the best plans, across all our benchmarks, 
and is best for most queries; Volcano-RU, which is cheaper, may be 
appropriate for inexpensive queries. 

Our multi-query optimization algorithms were partially prototyped on 
Microsoft SQL Server in summer '99, and are currently being evaluated
by Microsoft for possible inclusion in SQL Server.

In conclusion, we believe we have laid the groundwork for practical use of
multi-query optimization, and {\em multi-query optimization will form a 
critical part of all query optimizers in the future}.

\vspace{1mm}
\noindent{\large\bf Acknowledgments}:
{\small 
This work was supported in part by a grant from Engage Technologies/Redbrick
Systems.
Part of the work of Prasan Roy was supported by an IBM Fellowship.
We wish to thank K. Sriravi, who helped build the basic Volcano-based
optimizer, and Dan Jaye and Ashok Sawe for motivating this work through 
the Engage.Fusion project.  We also thank Krithi Ramamritham
and Sridhar Ramaswamy for providing feedback on the paper,
and Paul Larson both for feedback on the paper, and for 
inviting Prasan Roy to participate in prototyping our algorithms 
on SQL Server at Microsoft.
}